\RequirePackage{silence} %
\documentclass[ twoside,openright,titlepage,numbers=noenddot,%
                headinclude,footinclude,cleardoublepage=empty,abstract=on,
                BCOR=5mm,paper=a4,fontsize=11pt
                ]{scrreprt}

\PassOptionsToPackage{utf8}{inputenc}
  \usepackage{inputenc}

\PassOptionsToPackage{T1}{fontenc} %
  \usepackage{fontenc}

\usepackage{placeins}

\PassOptionsToPackage{
  drafting=false,    %
  tocaligned=false, %
  dottedtoc=false,  %
  eulerchapternumbers=true, %
  linedheaders=false,       %
  floatperchapter=true,     %
  eulermath=false,  %
  beramono=true,    %
  palatino=true,    %
  style=classicthesis %
}{classicthesis}

\newcommand{\myTitle}{Non-equilibrium dynamics of a binary solvent around heated colloidal particles\xspace}
\newcommand{\mySubtitle}{Bachelor Thesis\xspace}

\newcommand{\myName}{Moritz Wilke\xspace}

\newcommand{\myFaculty}{Max-Planck-Institute\xspace}

\newcommand{\myUni}{University of Stuttgart\xspace}
\newcommand{\myLocation}{Stuttgart\xspace}
\newcommand{\myTime}{August 16th, 2018\xspace}
\newcommand{\myVersion}{\classicthesis}

\providecommand{\mLyX}{L\kern-.1667em\lower.25em\hbox{Y}\kern-.125emX\@}

\PassOptionsToPackage{ngerman,american}{babel} %
    \usepackage{babel}

\usepackage{csquotes}
\PassOptionsToPackage{%
  backend=bibtex8,bibencoding=ascii,%
   sorting=nyt, %
}{biblatex}
    \usepackage{biblatex}

\PassOptionsToPackage{fleqn}{amsmath}       %
  \usepackage{amsmath}

\usepackage{graphicx} %
\usepackage{scrhack} %
\usepackage{xspace} %
\PassOptionsToPackage{printonlyused,smaller}{acronym}
  \usepackage{acronym} %
\usepackage{tabularx} %
\usepackage{subfig}
\usepackage{listings}
\usepackage{classicthesis}

\hypersetup{%
  colorlinks=true, linktocpage=true, pdfstartpage=3, pdfstartview=FitV,%
  breaklinks=true, pageanchor=true,%
  pdfpagemode=UseNone, %
  plainpages=false, bookmarksnumbered, bookmarksopen=true, bookmarksopenlevel=1,%
  hypertexnames=true, pdfhighlight=/O,%
  urlcolor=CTurl, linkcolor=CTlink, citecolor=CTcitation, %
  pdftitle={\myTitle},%
  pdfauthor={\textcopyright\ \myName, \myUni, \myFaculty},%
  pdfsubject={},%
  pdfkeywords={},%
  pdfcreator={pdfLaTeX},%
  pdfproducer={LaTeX with hyperref and classicthesis}%
}

\makeatletter
\@ifpackageloaded{babel}%
  {%
    \addto\extrasamerican{%
    }%
    \addto\extrasngerman{%
    }%
    }{\relax}
\makeatother

\usepackage{tikz}
\usetikzlibrary{arrows}
\usetikzlibrary{patterns}
\usepackage{braket}
\usepackage{epstopdf}
\usepackage{layouts}
\usepackage{graphicx}%
\usepackage{dcolumn}%
\usepackage{bm}%
\usepackage{epsfig}
\usepackage {mathrsfs}
\usepackage {fix-cm}
\DeclareMathAlphabet{\mathpzc}{OT1}{pzc}{m}{it}

\usepackage{indentfirst} 
\usepackage[capitalize]{cleveref}
\crefrangelabelformat{equation}{(#3#1#4)$-$(#5#2#6)}
\usepackage[left=1.5in,right=1.5in,top=1in,bottom=1in]{geometry}

\begin{document}
\frenchspacing
\raggedbottom
\selectlanguage{american} %
\pagenumbering{roman}
\pagestyle{plain}
\begin{titlepage}
    \begin{addmargin}[-3cm]{-3cm}
    \begin{center}
        \large

        \hfill

        \vfill

        \begingroup
            \color{CTtitle}\spacedallcaps{\myTitle} \\ \bigskip
        \endgroup
        
        \spacedallcaps{\myName}

        \vfill

        \includegraphics[width=5cm]{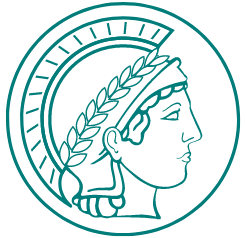} \\ \medskip

        \mySubtitle \\ \medskip
		\vspace{1.5cm}
        Max-Planck Institute for Intelligent Systems \\
        \myUni \\ \bigskip
        Supervisor: Dr. Sutapa Roy\xspace \\
        Examiner: Priv.-Doz. Dr. Markus Bier\xspace \\ \bigskip

        Thesis submitted on: \myTime\ %

        \vfill

    \end{center}
  \end{addmargin}
\end{titlepage}

\thispagestyle{empty}

\hfill

\vfill

\noindent\myName: \textit{\myTitle,} \mySubtitle, %
\textcopyright\ \myTime

\cleardoublepage
\pdfbookmark[1]{Abstract}{Abstract}
\begingroup
\let\clearpage\relax
\let\cleardoublepage\relax
\let\cleardoublepage\relax

\chapter*{Abstract}
Using numerical simulations, we study the non-equilibrium coarsening 
dynamics of a binary solvent around spherical colloids in the presence of 
a temperature gradient. The coarsening dynamics following a 
temperature quench is studied by solving the 
coupled modified Cahn-Hilliard-Cook equation and the heat diffusion 
equation, which describe the concentration profile and the temperature field, respectively. For the temperature field 
we apply a suitable boundary condition. We observe the formation of 
circular layers of different phases around the colloid whereas away 
from the colloid patterns of spinodal decomposition persist. Additionally, 
we investigate the dependence of the pattern formation on the quench 
temperature. Our simulation mimics an experimental system where the 
colloid is heated by laser illumination. Note that we look at the 
cooling of a solvent with an upper critical temperature, whereas the 
experimental analogue is the laser-heating of a solvent with a lower 
critical temperature. We also study a two colloid system. Here, we 
observe that a bridge of one phase forms connecting the two colloids. 
Also, we study the force acting on the colloids that is generated by 
the chemical potential gradient.

\begin{otherlanguage}{ngerman}
\pdfbookmark[1]{Zusammenfassung}{Zusammenfassung}
\chapter*{Zusammenfassung}
Mit Hilfe numerischer Simulationen untersuchen wir die nicht-gleichgewichts Dynamik der Phasentrennung in Anwesenheit eines Temperaturgradienten an sph"arischen kollodialen Teilchen.
Die nach einem Abschreckvorgang folgende Phasentrennung wird unter Verwendung der gekoppelten Cahn-Hilliard-Cook Gleichung und der W"armeleitungsgleichung, die das Konzentrationsfeld und das Temperaturfeld beschreiben, untersucht. F"ur das Temperaturfeld benutzen wir eine passende Randbedingung. Wir beobachten, dass sich kreisf"ormige Schichten verschiedener Phasen um das Kolloid herum bilden, wohingegen entfernt vom Kolloid Muster erhalten bleiben, die an spinodale Entmischung erinnern. Zus"atzlich untersuchen wir die Abh"angigkeit der Formation der Muster von der Temperatur, auf die das Kolloid abgeschreckt wird. Unsere Simulation imitiert eine experimentelle Realisierung der Strukturbildung um in einem L"osungsmittel suspendierte Kolloide, die mit einem Laser erhitzt werden. Es ist anzumerken, dass der hier behandelte Abschreckvorgang eines L"osungsmittels mit einer oberen kritischen Temperatur einem experimentellen Erhitzen eines L"osungsmittels mit unterer kritischen Temperatur entspricht. Zus"atzlich untersuchen wir ein zwei-Kolloid System. In diesem Fall formt sich eine Fl"ussigkeitsbr"ucke, die beide Kolloide miteinander verbindet. Au"serdem untersuchen wir die Kraft, die aus dem Gradienten des Feldes des chemischen Potentials r"uhrt und auf die Kolloide wirkt.
\end{otherlanguage}

\endgroup

\vfill
\cleardoublepage
\pagestyle{scrheadings}
\pdfbookmark[1]{\contentsname}{tableofcontents}
\setcounter{tocdepth}{2} %
\setcounter{secnumdepth}{3} %
\manualmark
\markboth{\spacedlowsmallcaps{\contentsname}}{\spacedlowsmallcaps{\contentsname}}
\tableofcontents
\automark[section]{chapter}
\renewcommand{\chaptermark}[1]{\markboth{\spacedlowsmallcaps{#1}}{\spacedlowsmallcaps{#1}}}
\renewcommand{\sectionmark}[1]{\markright{\textsc{\thesection}\enspace\spacedlowsmallcaps{#1}}}
\clearpage
\cleardoublepage
\pagestyle{scrheadings}
\pagenumbering{arabic}
\cleardoublepage
\chapter{Introduction}\label{ch:introduction}
Phase separation of binary fluids is a process that everybody 
has seen before in their daily lives. A good example may be an oil-water 
mixture. At room temperature, the shaking of a salad mixture containing oil 
and water generates many small oil droplets getting separated out as they 
do not like to mix with water. Understanding the dynamics of such phase 
separation processes is a subject of great interest. 
They also hold important applications in oil and pharmaceutical industries, 
e.g., extraction of oil and natural gases from rocks, stability of foams, 
etc. John W. Cahn and John E. Hilliard devoted themselves to this 
non-equilibrium process of phase separation and derived in 1958 the 
Cahn-Hilliard equation \cite{binderpuribook}. It describes the time 
evolution of the local composition for spontaneous phase separation 
of a binary fluid.

Consider a homogeneous binary liquid mixture (A+B) above its 
critical temperature $T_c$. When we suddenly quench it below $T_c$, 
the binary liquid now lies in a non-equilibrium state and it separates 
into A-rich and B-rich domains which grow in size over time. The kinetics 
of this process is known as the \textit{phase separation dynamics} or 
\textit{coarsening}. In this thesis we consider only fluids with an 
upper critical temperature. That means phase separation only occurs 
when the temperature is below the critical temperature $T_\text{c}$. 
Coarsening processes were studied in detail for bulk systems 
\cite{binderpuribook}. It is of great interest to study the phase 
separation with effects of a surface \cite{dasprl} to find how phase 
separation processes change as compared to bulk.

We simulate a system where a single particle with a preference 
for one of the two components of the fluid is placed in the binary 
solvent. A lot of studies were devoted to understanding the phase 
separation phenomenons in for slit geometry. It is new 
\cite{roypre,roysoftmatter} to investigate them around spherical 
colloidal particles. We are interested in the coarsening of a binary 
solvent around colloids in the presence of a time-dependent 
\textit{temperature gradient}. Such systems are experimentally relevant 
for laser-heated colloids in a binary fluid, which have been considered 
in recent years \cite{bechinger}. It was demonstrated that when Janus 
particles suspended in a critical binary liquid were illuminated by light 
it generates a concentration gradient around the colloid \cite{bechinger}. 
This local demixing of the binary liquid at early time was hardly 
investigated \cite{roypre,roysoftmatter,royruben}. This non-equilibrium 
dynamics is interesting and complex because of the surface effects of 
the particle, time-dependent temperature gradient and local phase 
separation.

So far, coarsening dynamics around heated colloids was studied 
for a single colloid only \cite{roypre, roysoftmatter}. 
In this thesis, we look at the coarsening dynamics around 
two colloids suspended in a near-critical binary solvent. It is not 
easy to guess how the presence of the second colloid will change the 
kinetics. This is complicated because of the temperature gradient 
coupled to the concentration field. 
The explored quantities are the time-dependent temperature field 
and the time-dependent order parameter 
field. They are described by the modified Cahn-Hilliard-Cook equation 
and the heat diffusion equation. 
We also study the coarsening properties around a single colloid with a 
temperature boundary condition (b.c.) 
which is different from what was considered in previous works \cite{roypre, roysoftmatter}. 
Different b.c.s determine the temperature field and thus, they can control 
the local structure formation. We also study the 
forces acting on the colloid that are generated due to the gradient of the chemical potential field. 
For a single colloid, this force is zero and the colloid does not move. But, when two 
colloids are suspended in a solvent, there may be a non-zero force.

It is important to mention that we deal with a binary solvent 
that has an upper critical point. Therefore, to enter the two-phase 
state of the solvent, we must apply a temperature quench below $T_c$. 
The experimental analogue is the heating of a binary solvent with an 
upper critical point because it is much more practical and accessible 
through laser-illumination.

Rest of the thesis is structured as follows: In Sec. \ref{ch:theory} 
we explain the basic theoretical backgrounds to understand the phase 
separation dynamics of a binary solvent and how to describe it. In Sec. 
\ref{model} we derive the model of our system and mention the numerical 
methods we use to simulate it. Complementary, we lead through necessary 
calculations in the appendix (Sec. \ref{ch:appendix}). Section \ref{ch:results} 
presents and discusses our findings for a single colloid and a two 
colloid system. Lastly, Sec. \ref{ch:summary} gives a short overview 
about the content of the thesis and mentions possible subsequent studies.

\cleardoublepage
\chapter{Theory}\label{ch:theory}
\section{Phase separation dynamics}

\FloatBarrier
\begin{figure}[h]
\begin{center}
\begin{tikzpicture}[
	scale=1.4,
    axis/.style={line width=0.6mm, ->, >=stealth'},
    every node/.style={color=black}]

	\draw[axis] (-3,0) -- coordinate (x axis) (3.3,0);
	\node[below=0] at (0,-0.4){Concentration $c_\text{A} = \frac{N_\text{A}}{N_\text{A}+N_\text{B}}$};
	\draw[axis] (-3,0) -- coordinate (y axis) (-3,5.5);
	\node[rotate = 90, above=15] at (-3.15,2.75){Temperature $T$};

	\node (critical quench) at (0,4.6) {Critical quench}; 
	\draw[very thick, ->, >=stealth'] (critical quench) -- (0,1.5);
	\node (off-critical quench) at (-1.7,4) {Off-critical quench}; 
	\draw[very thick, ->, >=stealth'] (off-critical quench) -- (-1.7,2.4);
	\draw[thick, ->, >=stealth', draw=none] (0,1.5) -- (-2.88675,1.5) node[pos=0.35,fill=white] {Coarsening};
	\draw [color=black!40!green] (2.41455,1.5) circle [radius=0.8pt];
	\draw [color=black!40!green] (-2.41455,1.5) circle [radius=0.8pt];
	\draw [color=black!40!green] (-1.92684,2.4) circle [radius=0.8pt];
	\draw [color=black!40!green] (1.92684,2.4) circle [radius=0.8pt];

    \draw[domain=0:2.94782, blue=20!] plot(\x,{3.33-0.13*\x*\x*\x}); %
	\draw[domain=-2.94782:0, blue=20!] plot(\x,{3.33+0.13*\x*\x*\x}); %
	\draw[dashed, domain=-2.70589:2.70589, orange] plot(\x,{-0.5*\x*\x+0.00617284*\x*\x*\x*\x+3.33}); %

	\filldraw [blue=20!] (0,3.33) circle [radius=1.5pt] node (critical) [above=15, right=10, draw, rectangle] {Critical point};
	\draw (0,3.33) -- (critical);

	\node at (0,5.15) {MIXED REGION};
	\node at (0.1,0.75) {DEMIXED REGION};
	\node[fill=white] at (2,2.716672) {Binodal};
	\node[fill=white] at (1.5,1.78) {Spinodal};

	\draw (-2.95,3.33) -- (-3.05,3.33) node[anchor=east] {$T_c$};
	\draw (0,0.05) -- (0,-0.05) node[anchor=north] {$c_\text{c}$};

\end{tikzpicture}
\caption{Schematic phase diagram of a binary liquid mixture with 
components A and B in the temperature ($T$) -- concentration ($c_\text{A}$) 
plane. $T_c$ and $c_c$ correspond to the critical temperature and the critical 
concentration of the liquid, respectively. The solid blue line denotes 
the binodal or co-existence curve and the dashed orange line the spinodal. 
The left side of the binodal refers to A-rich phases, whereas the right side 
refers to B-rich phases. For $T>T_c$ a liquid at critical concentration is 
in a homogeneously mixed state and for $T<T_c$ it is in a demixed state. 
A temperature quench from a high temperature ($>T_c$) to a 
temperature $T<T_c$ at the critical composition leads to coarsening 
via spinodal decomposition (see main text for detail). On the other hand, 
an off-critical quench inside the region between the binodal and spinodal 
curves leads to coarsening via the nucleation of spherical droplets.} 
\label{fig:phasediagram}
\end{center}
\end{figure}
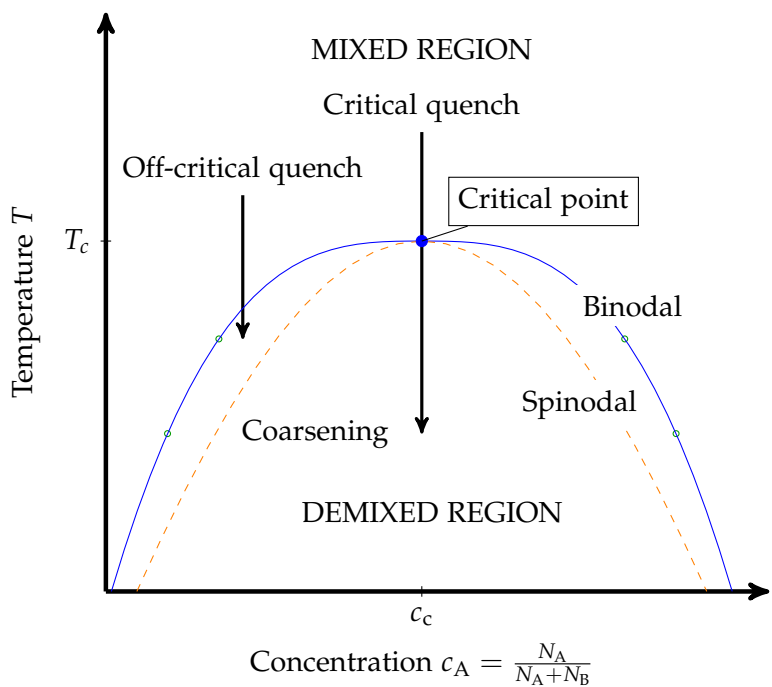
\FloatBarrier

Phase separation is the process of unmixing of different phases of a system. 
Let us consider a binary liquid mixture having A and B types of particles. 
To explain phase separation, let us consider the phase diagram of this liquid mixture in 
the temperature ($T$) -- concentration ($c_\text{A}$) 
plane, as shown in Fig. \ref{fig:phasediagram}. Here, the concentration of species A is defined as 
$c_A=N_A/(N_A+N_B)$; $N_A$ is the number of particles of species $A$. 
The solid blue curve is the co-existence curve or the `binodal' along which 
the two phases can co-exist with each other in equilibrium. 
Specifically, the right branch of it corresponds to the A-rich phase and the left branch to the 
B-rich phase. $T_c$ and $c_\text{c}$ are the values of the temperature and the concentration at the 
critical point, respectively. We will explain the broken curve later. 
Above $T_c$, the equilibrium state is the homogeneously mixed state and below $T_c$, it is the 
demixed state. 

Now, when an initially homogeneous binary liquid mixture at a temperature $T>T_c$ 
is suddenly quenched inside the 
binodal, the system falls out of equilibrium and then it moves towards the new 
equilibrium state which is the phase separated state with equilibrium composition 
values marked by green circles.
However, this does not happen immediately. During the intermediate time the system goes through
a complex dynamics \cite{binderpuribook} which is known as the \textit{phase separation dynamics} or the 
\textit{coarsening dynamics}. During such a process, domains of A- and B- rich phases 
form and they grow with time.
 
The `spinodal' denotes the boundary between the region where the 
fluid will phase separate via spinodal decomposition, which is explained 
in the following text, and the region where the fluid is metastable. 
A metastable binary fluid stays in equilibrium for small concentration 
fluctuations but will phase separate for larger ones.

Kinetics of the phase separation process depends on the concentration of the liquid during the quench. If it is at its `critical' concentration, percolating domains form \cite{royjcp}. This mechanism is known as 
the \textit{spinodal decomposition} \cite{binderpuribook}. 
Typically, the average domain size, $\ell(t)$, during phase 
separation grows in a power-law with a growth exponent $\alpha$, $\ell(t) \sim t^\alpha$. 
For spinodal decomposition, at early time $\alpha=1/3$ which corresponds to the diffusive dynamics. 
At very late times, the growth exponent changes. However, in this 
thesis we will consider the diffusive dynamics only. On the other hand, when the fluid is quenched at an `off-critical' concentration 
inside the metastable region in between the 
binodal and the spinodal, phase separation occurs following \textit{nucleation} \cite{binderpuribook} of droplets. 
\newpage
\section{Ginzburg-Landau free energy functional}

\FloatBarrier
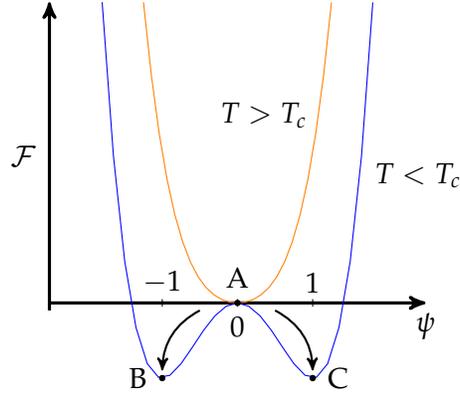
\begin{figure}[h]
\begin{center}
\begin{tikzpicture}[
	scale=1,
    axis/.style={very thick, ->, >=stealth'},
    every node/.style={color=black}]
    
	\draw[very thick, >=stealth', ->] (-2.5,0) -- coordinate (x axis) (2.5,0);
	\node[below=0] at (2.5,0){$\psi$};
	\draw (-1,0.05) -- (-1,-0.05);
	\node[above] at (-1,0) {$-1$};
	\draw (0,0.05) -- (0,-0.05) node[anchor=north] {$0$};
	\draw (1,0.05) -- (1,-0.05); %
	\node[above] at (1,0) {$1$};
	\draw[axis] (-2.5,-0.1) -- coordinate (y axis) (-2.5,4);
	\node[rotate = 0, left=1.25] at (y axis){$\mathcal{F}$};
	\draw[domain=-1.2496:1.2496,variable=\x, orange] plot({\x},{\x*\x+\x*\x*\x*\x}); 
	\draw[domain=-1.7989:1.7989,variable=\x, blue=20!] plot({\x},{-2*\x*\x+\x*\x*\x*\x});
	\node[above=2] at (0,0) {A};
	\node[left=2] at (-1,-1) {B};
	\node[right=2] at (1,-1) {C};
	\node at (0.35,2.5) {$T>T_c$};
	\node at (2.4,1.7) {$T<T_c$};
	\filldraw [black] (0,0) circle [radius=1pt];
	\filldraw [black] (-1,-1) circle [radius=1pt];
	\filldraw [black] (1,-1) circle [radius=1pt];
	\draw[thick, ->, >=stealth'] (-0.5,-0.1) to [bend right] (-1,-0.9);
	\draw[thick, ->, >=stealth'] (0.5,-0.1) to [bend left] (1,-0.9);
\end{tikzpicture}%

\end{center}
\caption{Schematic plot of the Landau free energy $\cal F$ as a 
function of the order parameter $\psi$. The orange curve corresponds to 
the free energy above the critical temperature $T_c$ with the minimum A 
as the equilibrium state. The blue curve corresponds to temperatures below 
$T_c$ with minima B and C.}
\label{fig:landau}
\end{figure}
\FloatBarrier

Coarsening dynamics can be understood by considering the Landau free 
energy whose function form is given below:
 
\begin{equation} \label{landau}
{\cal F}(\psi, T)=V~ \Big [\frac{a}{2}\psi^2 + \frac{u}{4}\psi^4 \Big],
\end{equation}
where, $\psi$ is the order parameter we introduce as the concentration 
difference of the two species $\psi =  c_A - c_B$, $V$ the volume, $a$ and 
$u$ are phenomenological parameters; $a \propto (T-T_c)$, and $u>0$. 
For $T>T_c$, $a>0$ and vice-versa. This free energy is schematically shown 
in Fig. \ref{fig:landau}. The equilibrium state of a system corresponds to 
the minimum of the free energy. It can be seen that for $T>T_c$ there is 
only one minimum (point ``A'') at a order parameter value $\psi =0$ which 
is the fully mixed state. That means, for temperatures higher than $T_c$ 
the thermodynamically favourable state of a system is the mixed state. 
For $T<T_c$, there are two minima (``B'' and ``C''). The system then wants 
to be in a phase separated state with the order parameter values given by 
the minima. Thus, upon a sudden temperature 
quench from $T>T_c$ to a temperature inside the binodal curve, the system will fall out 
of equilibrium and try to move from point ``A'' to ``B'' (or ``C''). However, as mentioned before, 
it can not move immediately. The system goes through a non-equilibrium dynamics 
during which domains form and grow with time. 
 
A phase separating fluid is spatially non-uniform. To describe the 
behavior of a phase separating fluid, we need to allow the order parameter to be 
space dependent, $\psi(\vec r)$. The free energy is now a functional of $\psi(\vec r)$ and 
following Ginzburg, it takes up the functional form:
\begin{equation} \label{ginzburg-landau}
{\cal F}[\psi(\vec r), T]= \frac{k_B T_c}{v} \int d^3 {\vec r} ~ \Big[\frac{a}{2}\psi^2(\vec r) + 
\frac{u}{4}\psi^4(\vec r) + \frac{C}{2} (\nabla \psi)^2 \Big].
\end{equation}
This equation is known as the \textit{Ginzburg-Landau} free energy 
\cite{GLpaper}. Here, $C$ is a phenomenological parameter and 
$v$ is an elementary volume element.

\section{Cahn-Hilliard-Cook equation}\label{sec:chc}
During the phase separation of binary mixture (A+B), the concentration 
of A and B species is conserved and as a result the order parameter is also 
conserved. The dynamics of such a system is given by the so-called `Model B' 
dynamics \cite{binderpuribook}. The local conservation in this case requires 
a continuity equation
\begin{align} \label{eq:continuityeq}
\frac{\partial \psi(\vec{r},t)}{\partial t} = - \nabla \cdot \vec{j}(\vec{r},t),
\end{align}
where, the current, $\vec{j}(\vec{r},t)$, is proportional to the gradient of 
the chemical potential 
$\mu$ as
\begin{align} \label{eq:flux}
\vec{j}(\vec{r},t) = -M \nabla \mu(\vec{r},t).
\end{align}
$M$ is the mobility. The chemical potential $\mu$ can be obtained as 
\begin{align} \label{eq:chempot}
\mu(\vec{r},t) = \frac{\delta \mathcal{F}[\psi]}{\delta \psi(\vec{r},t)}.
\end{align}
Combining Eqs. (\ref{eq:continuityeq}), (\ref{eq:flux}), and (\ref{eq:chempot}), 
one obtains
\begin{align} \label{eq:continuityeq2}
\frac{\partial \psi(\vec{r},t)}{\partial t} = M \nabla^2 \frac{\delta \mathcal{F}[\psi]}{\delta \psi(\vec{r},t)}.
\end{align}
Insertion of the Ginzburg-Landau free energy functional of Eq. (\ref{ginzburg-landau}) 
into the above equation yields 
\begin{align} \label{eq:CHC}
\frac{\partial \psi(\vec{r},t)}{\partial t} = \frac{M k_B T_c}{v}~ \nabla^2 \left[a \psi(\vec{r},t)
+u \psi(\vec{r},t)^3 -C\nabla^2 \psi(\vec{r},t) \right],
\end{align}
where, $a \propto (T-T_c)$. Equation (\ref{eq:CHC}) is known as the 
\textit{Cahn-Hilliard-Cook} (CHC) equation or ``model B'' equation. 

Note that the CHC equation in Eq. (\ref{eq:CHC}) describes a phase 
segregating system where the temperature is same everywhere in the system. 
Most of the literature is about simulating the process of coarsening 
after a temperature quench to a constant temperature of the whole fluid. 
For systems with a temperature gradient, i.e., when $T$ depends on position 
$T(\vec r)$, we need to modify Eq. (\ref{eq:CHC}). We will describe this 
later in the Sec. \ref{model} of the thesis.

\chapter{Model and Method}\label{model} %
Our model system consists of spherical colloid (3-d) suspended 
in a binary solvent with an upper critical temperature $T_c$. Initially, 
the colloid and the solvent are set to a temperature $T_0$ which is higher 
than the demixing critical temperature $T_c$ of the binary solvent, so 
that the solvent is in a mixed phase.
Next, at time $t=0$, the colloid(s) is(are) quenched down to a temperature $T_1$ below $T_c$. 
After this, a heat flow cools the solvent and a time-dependent temperature gradient 
is established in the system. 
As the solvent temperature cools below the demixing temperature $T_c$, the solvent starts to phase 
separate. We describe the non-equilibrium dynamics of the solvent after the quench with 
two fields: the temperature $T(\vec r,t)$ and the order parameter $\psi(\vec r,t)$ 
fields. The order parameter 
field is coupled to the temperature field. This coupling, together with the colloid surface effects, 
make the coarsening process more complex as compared to an instantaneous quench and coarsening in bulk. 

\section{Modified Cahn-Hilliard-Cook equation and boundary conditions}
Here we explain the time evolution of $T(\vec r,t)$ and $\psi (\vec r,t)$. 
The order parameter field is governed by the \textit{Cahn-Hilliard-Cook} 
equation (described before in Sec. \ref{sec:chc}) for the conserved order 
parameter. However, the original CHC equation in Eq. (\ref{eq:CHC}) is for a 
system where temperature is same everywhere in the system. In order to take 
into account the temperature gradient present in our system we modify the equation 
appropriately. We replace the phenomenological parameter $a$ by the temperature 
field $\tilde T(\vec r,t)={\cal A}(T(\vec r,t)-T_c)/T_c$ \cite{roypre}, where, 
$\cal A$ is a positive constant. The modified CHC equation then reads
\begin{align} \label{eq:CHCdim}
\frac{\partial \psi(\vec{r},t)}{\partial t} 
&= \frac{M k_B T_c}{v}~ \nabla^2 \left[ \tilde{T}(\vec{r},t) \psi(\vec{r},t) +u\psi(\vec{r},t)^3 -C \nabla^2 \psi(\vec{r},t) \right]
\end{align}

In order to consider thermal fluctuations, Gaussian white noise 
$\eta(\vec{r},t)$ is added to Eq. (\ref{eq:CHCdim}):
\begin{align} \label{eq:CHCnoise}
\frac{\partial \psi(\vec{r},t)}{\partial t} &= \frac{M k_B T_c}{v}~ \nabla^2 \left[ \tilde{T}(\vec{r},t) \psi(\vec{r},t) +u\psi(\vec{r},t)^3 -C \nabla^2 \psi(\vec{r},t) \right] \nonumber  \\
&\quad + \eta(\vec{r},t).
\end{align}
This Gaussian white noise has zero mean and satisfies the 
fluctuation-dissipation relation
\begin{align} \label{eq:whitenoiserelation}
\braket{\eta(\vec{r},t)\eta(\vec{r}',t')} = -\frac{2M}{v}k_B T(\vec{r}) \nabla^2 \delta(\vec{r}-\vec{r}')\delta(t-t').
\end{align}

Equation (\ref{eq:CHCnoise}) is made dimensionless by scaling the 
dimensional variables $r$, $t$, $\psi$, $\eta$ by scaling factors. This 
calculation is shown in Sec. \ref{sec:nondimensionalizationCHCandHD}. 
With the following rescaling factors (where variables without and with 
\textasciitilde \ are the dimensional and dimensionless ones, respectively), 
the dimensionless modified CHC equation becomes

\begin{align}
\psi_0 &= \left(\frac{|\tilde{T}_1|}{u}\right)^{1/2}, \label{eq:rescalingcoefficientPsi0} \\
r_0 &= \left(\frac{C}{u \psi_0^2}\right)^{1/2} = \left(\frac{C}{|\tilde{T}_1|}\right)^{1/2}, \label{eq:rescalingcoefficientr0} \\
t_0 &= \frac{\nu r_0^2}{M k_B T_c |\tilde{T}_1|} = \frac{\nu C}{M k_B T_c |\tilde{T}_1|^2},\\
\eta_0 &= \frac{\psi_0}{t_0} = \left(\frac{|\tilde{T}_1|}{u}\right)^{1/2}/\frac{\nu C}{M k_B T_c |\tilde{T}_1|^2},
\end{align} 
\begin{equation} \label{eq:CHC2}
\frac{\partial {\tilde \psi}(\tilde {\vec r}, \tilde t)}{\partial \tilde t} =
{\tilde \nabla}^2 \left[\frac{{\tilde T}(\tilde {\vec r}, \tilde t)}{{\tilde T}_1}{\tilde \psi}(\tilde {\vec r}, \tilde t) 
+{\tilde \psi}(\tilde {\vec r}, \tilde t)^3 
-{\tilde \nabla}^2 {\tilde \psi}(\tilde {\vec r}, \tilde t) \right] 
+ {\tilde \eta}(\tilde {\vec r}, \tilde t).
\end{equation}

The time evolution of the temperature field $\tilde T$ is described by 
the \textbf{heat diffusion equation}
\begin{align} \label{eq:HDdim}
\frac{\partial \tilde{T}(\vec{r},t)}{\partial t} = D_\text{th} \nabla^2 \tilde{T}(\vec{r},t),
\end{align}
whose dimensionless form is given below (for the derivation see Sec. 
\ref{sec:nondimensionalizationCHCandHD}):
\begin{align} \label{eq:HD}
\frac{\partial \tilde{T}(\tilde {\vec r}, \tilde t)}{\partial {\tilde t}} = 
\mathcal{D} \tilde{\nabla}^2 {\tilde T}(\tilde {\vec r},\tilde t),
\end{align}
where, $\mathcal{D} = D_\text{th}/(|\tilde T_1| D_\text{m})$, $D_\text{th}$ 
is the thermal diffusivity of the binary solvent and $D_\text{m}=M(k_B T_c/v)$ 
is the interdiffusivity of the solvent at $T_c$. 

Equation (\ref{eq:HD}) is solved subject to boundary conditions (b.c.)
\begin{align} 
\tilde{T}(\vec{r},t)|_{\mathcal{C}} = \tilde{T}_1  \label{eq:boundarytemp} 
\end{align}
where, $\mathcal{C}$ refers to the surface of the colloid. To maintain the 
temperature at the outer edges of the simulation box fixed to $\tilde T_0$, we 
use the following b.c. 
\begin{align} 
\frac{\partial \tilde T(\tilde{\vec r},\tilde t)}{\partial \tilde t}= \mathcal{D} \tilde{\nabla}^2 {\tilde T}(\tilde {\vec r},\tilde t)
-c(\tilde T(\tilde{\vec r},\tilde t) - \tilde T_0), \label{eq:boundarytemp2}
\end{align}
here, the sink term $-c(\tilde T(\tilde{\vec r},\tilde t) - \tilde T_0)$ mimics heat dissipation. 
We choose $c=0.001$ which ensures that $\tilde T$ at the outer boundaries is at $\tilde T_0$.

In this thesis we consider a surface that attracts one of the two components 
of the fluid. To describe this preference, we consider a surface free energy 
term \cite{roypre}
\begin{align}
\mathcal{F}_S = \left [\frac{1}{2} \alpha \int_{\mathcal{S}} \psi(\vec{r},t)^2 dS 
- h \int_{\mathcal{S}} \psi(\vec{r},t) dS \right], 
\end{align}
which we add to $\mathcal{F}$. Here, $\mathcal{S}$ is the surface of the 
colloid, $\alpha$ is the surface enhancement parameter and $h$ is a symmetry-breaking 
surface field. The state where the free energy is lowest is the thermodynamically 
favorable state. That is why $\psi$ is negative for positive $\alpha$ and the other 
way around. Thus $\alpha$ determines which phase is present near the colloid. This 
gives rise to an additional b.c. 
\begin{align} \label{eq:robin}
\hat{\vec{n}} \cdot \tilde {\nabla} \tilde {\psi}(\tilde {\vec{r}},\tilde t)|_{\mathcal{S}} = 
-\tilde{\alpha} \tilde{\psi}(\tilde {\vec{r}},\tilde t)|_{\mathcal{S}} + \tilde{h} 
\end{align}
which is the so-called \textbf{Robin} boundary condition, where $\hat{\vec{n}}$ 
is a unit vector pointing into the colloid normal to the surface of the colloid. 
Here, the substitutions for non-dimensionalization are 
$\alpha = \left( C/|{\tilde{T}}_1| \right)^{-1/2}\tilde{\alpha}$ and 
$h = [|{\tilde{T}}_1|/(uC)^{1/2}] \tilde{h}$, that are calculated in 
Sec. \ref{sec:nondimensionalizationRobin}. Another necessary boundary condition is 
that there is no particle flux normal to the surface of the colloid allowed:
{\begin{align}\label{eq:noflux}
(\hat{\vec{n}} \cdot \nabla \mu(\vec{r},t))|_{\mathcal{S}} = (\hat{\vec{n}} \cdot \nabla \delta \mathcal{F}[\psi]/\delta \psi(\vec{r},t))|_{\mathcal{S}}=0.
\end{align}

\vspace{0.5cm}

\section{Numerical procedure} \label{sec:methods}
For the one-colloid system, a spherical particle of radius $R$ is kept fixed 
at the center of a cubic simulation box. To start with, every simulation grid point 
is assigned an initial temperature value $\tilde{T}_0 = 0.2$ and the grid points 
corresponding to the solvent are given an uniformly randomly generated order parameter 
value in the range of $[-1,1]$ with the condition that the spatially averaged order 
parameter is $\psi_0 = 0.2$. In Fig. \ref{fig:methods}, the setup is visualized for 
a single colloid system. At time $t=0$, the colloid grid points are set to $\tilde{T}_1= -1$ 
which mimics a thermal quench. The time evolution of the system is next described by solving 
numerically Eqs. (\ref{eq:CHC2}) and (\ref{eq:HD}) using the Euler's method \cite{euler}. 
For the numerical purpose, the discrete form of the gradient and the Laplacian are used. 
For a function $f(x,y,z)$ the discrete gradient is
\begin{align}
\nabla f(x,y,z) &\approx [f(x+1,y,z) - f(x,y,z) ]\vec{e}_x \nonumber \\
&\quad + [f(x,y+1,z) - f(x,y,z)]\vec{e}_y \nonumber \\
&\quad + [f(x,y,z+1) - f(x,y,z)]\vec{e}_z
\end{align}
and the discrete Laplacian is
\begin{align}
\nabla^2 f(x,y,z) &\approx f(x+1,y,z) + f(x,y+1,z) + f(x,y,z+1) \nonumber \\
 &\quad + f(x-1,y,z) + f(x,y-1,z) \nonumber \\
 &\quad + f(x,y,z-1) - 6 \cdot f(x,y,z).
\end{align}

\FloatBarrier
\begin{figure}[bth]
\begin{center}
\begin{tikzpicture}[
	scale=1.3,
    axis/.style={very thick, ->, >=stealth'},
    every node/.style={color=black}]
    
	\draw[color=orange] (0,0) rectangle (5,5);
	\node at (2.5,2.5) [circle,draw=blue!50,fill=blue!20, minimum size=2cm]{};
	\node at (2.5,2.5) {$\tilde{T}_1 = -1$};
	\draw (5,4.3) -- (5.4,4.3);
	\node at (6.4,4.3)  {$\tilde{T} = \tilde{T}_0 = 0.2$};
	\draw[very thick, <->, >=stealth'] (0,-0.2) -- (5,-0.2);
	\node at (2.5,-0.2) [below=1]  {$L$};
	\draw[very thick, <->, >=stealth'] (-0.2,0) -- (-0.2,5);
	\node at (-0.2,2.5) [left=1]  {$L$};

\end{tikzpicture}
\end{center}
\caption{Simulation-setup of a one-colloid system in a binary solvent shown in the midplane $(x,y)$ of the system ($z = L/2$). 
The colloid is drawn blue and is placed in the center of the box. It is set to a quenched temperature 
$\tilde{T}_1=-1$. Outside the colloid, each grid point is assigned an initial order parameter value from an 
uniformly generated random number distribution in the range $[-1:1]$, such that the mean order parameter $\psi_0=0$. 
Temperature at the outer edges of the simulation box are fixed to $\tilde T=\tilde{T}_0=0.2$.} 
\label{fig:methods}
\end{figure}
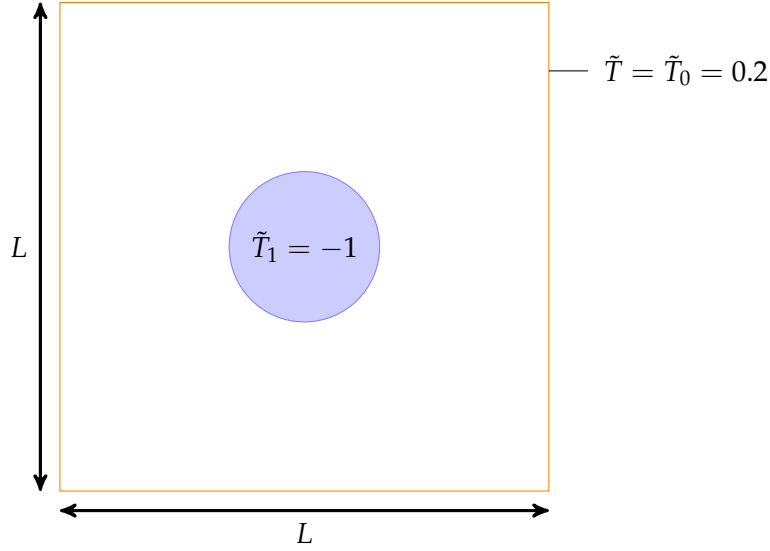
\FloatBarrier

The boundary conditions used are already mentioned before (Eqs. (\ref{eq:boundarytemp}), (\ref{eq:boundarytemp2}), 
(\ref{eq:robin}), and (\ref{eq:noflux})). We apply \textbf{periodic boundary conditions} on the outer edges of the simulation box. This is done to 
avoid any surface effects on the outer edges. 
For lattice points $[x(0), ... x(i), ... x(N)]$ on the x axis of a box with the length $x(N)$ 
the periodic boundary conditions are applied as follows:
\begin{align}
x(N+1) &= x(0) \nonumber \\
x(-1) &= x(N) \nonumber .
\end{align}}

For a two colloid system, we keep two spherical particles, each of radius $R$ in the cubic 
simulation box of side length $L$. The process of initial configuration generation is the same as 
for a single colloid system. At time $t=0$, both colloids are quenched to temperature $\tilde T_1=-1$. 
Outer edges of the simulation box are supplied with the periodic boundary condition. The b.c. for 
no flux in Eq. (\ref{eq:noflux}) is applied on both colloids.

All results were obtained for the parameters $\mathcal{D} = 50$, $\tilde{\alpha}=0.5$, $\tilde{h}=1$, and 
time step $\text{d}{\tilde t} = 0.001$. Also, all results presented in the 
thesis have been prepared using \textbf{Gnuplot} and \textbf{Xmgrace}.

\chapter{Results and discussion}\label{ch:results} %
Below we first present results for a single colloid system. 

\section{One colloid system}

We set up the simulation with a box size $L/r_0 = \tilde{L} = 100$ 
and the initial conditions as in Fig. \ref{fig:methods}, 
with $\tilde{T}_0 =0.2$, for a centered single colloid of radius $R/r_0 = \tilde{R} =10$. 
Note that $\mathcal{D} = 50$, $\tilde{\alpha}=0.5$, $\tilde{h}=1$ and $\text{d}\tilde{t} = 0.001$, 
as mentioned in the Sec. \ref{sec:methods}. The evolution of the system is then 
numerically calculated with the dimensionless heat diffusion equation (\ref{eq:HD}) and the 
coupled dimensionless Cahn-Hilliard-Cook equation (\ref{eq:CHC2}). 
\vspace{0.2cm}

\FloatBarrier
\begin{figure}[bth]
\centering
\includegraphics[scale=1]{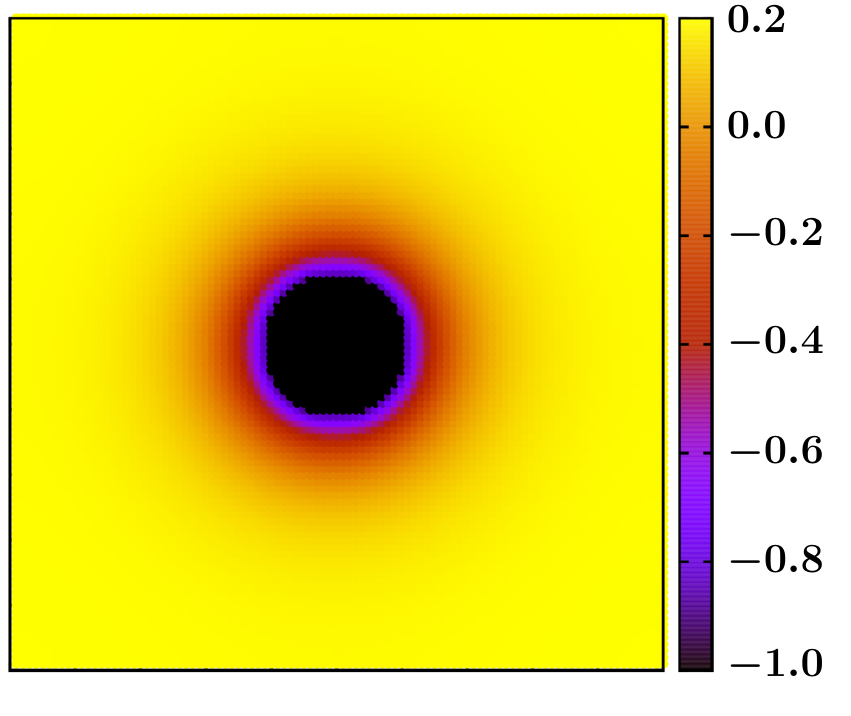}
\caption{Temperature profile of the one colloid system in the midplane ($z= L/2$), at an early time 
$\tilde{t}=10$. This time is 
during the non-equilibrium coarsening process after a temperature quench of the colloid from 
an initial temperature $\tilde T_0=0.2$ to $\tilde T_1=-1$. Initially, the solvent was also at $\tilde T_0=0.2$. 
After a quench, the surrounding solvent gets cooled due to heat flow. The outer boundaries of the simulation box 
are maintained at the initial temperature $0.2$ via appropriate temperature boundary condition. 
A large temperature gradient is established in the system.}
\label{fig:1colloid_temp}
\end{figure}
\FloatBarrier

In Fig. \ref{fig:1colloid_temp}, we show the temperature profile in the midplane 
($z= L/2$) at an early time $\tilde{t}=10$. 
The colloid is quenched to a temperature $\tilde T_1=-1$.
A large temperature gradient is present in the colloid's surrounding whereas the temperature close to
the boundaries of the box is near the initial value ${\tilde T}_0 = 0.2$. 
Although this is at an early time, the temperature 
field nearly has reached a steady state which is why there are no plots for different times necessary.

\begin{figure}[bth]
\centering
\includegraphics[scale=1]{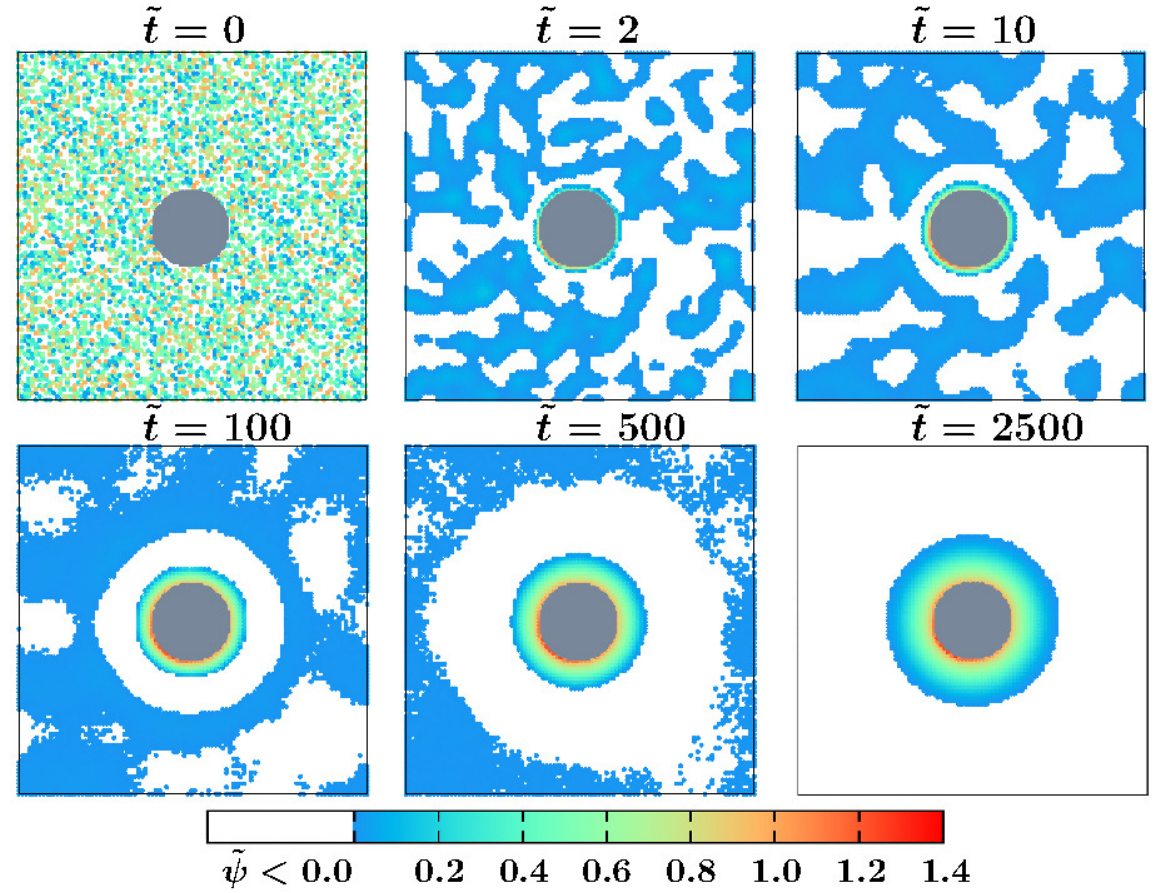}
\caption{Snapshots of temperature-gradient 
induced coarsening of a binary solvent around a spherical 
colloidal particle at six different times. 
Results correspond to $\tilde{L}=100$, $\tilde{R}=10$, 
$\tilde{\psi}_0=0$, $\tilde{T}_0=0.2$ and $\tilde{T}_1=-1$. 
The color bar denotes the value of the order parameter. 
Away from the colloid spinodal-like patterns persist over time. Close to the colloid the A phase 
($\tilde{\psi}(\tilde{\vec r},\tilde{t}) > 0$) accumulates as a 
layer around the colloid. As time progresses, 
the layer broadens and a layer of the B phase ($\tilde{\psi}(\tilde{\vec r},\tilde{t}) < 0$) forms next to it. The B phase broadens into the bulk to a large extend.}
\label{fig:1colloid_snapshot_T-1}
\end{figure}

Figure \ref{fig:1colloid_snapshot_T-1} shows the evolution of the order parameter 
in the midplane of the system at six different times $\tilde{t}=0$, $\tilde{t}=2$, $\tilde{t}=10$, $\tilde{t}=100$, $\tilde{t}=500$ and $\tilde{t}=2500$. 
At very early time $\tilde{t}=2$, the A phase, which we call the phase with 
$\tilde{\psi}(\tilde{\vec r},\tilde{t}) > 0$, already accumulates 
around the colloid because of the colloid's preference for the component A. 
Away from the colloid, a pattern comparable to spinodal decomposition is present. 
With time, the spinodal decomposition gets more prominent (see $\tilde {t}=10$). 
Also as time progresses, the phases merge into circular layers around the colloid. 
The A phase grows around the colloid with increasing values of the order parameter surrounded 
by a layer of the B phase ($\tilde{\psi}(\tilde{\vec r},\tilde{t}) < 0$). 
Away from the colloid a spinodal-like pattern preserves. 
Finally, at very late time, an almost circular very thick layer of A phase forms around the colloid 
(as seen at $\tilde t=2500$). 
\vspace{0.5cm}

\FloatBarrier
\FloatBarrier

\begin{figure}[bth]
\centering
\includegraphics[scale=1]{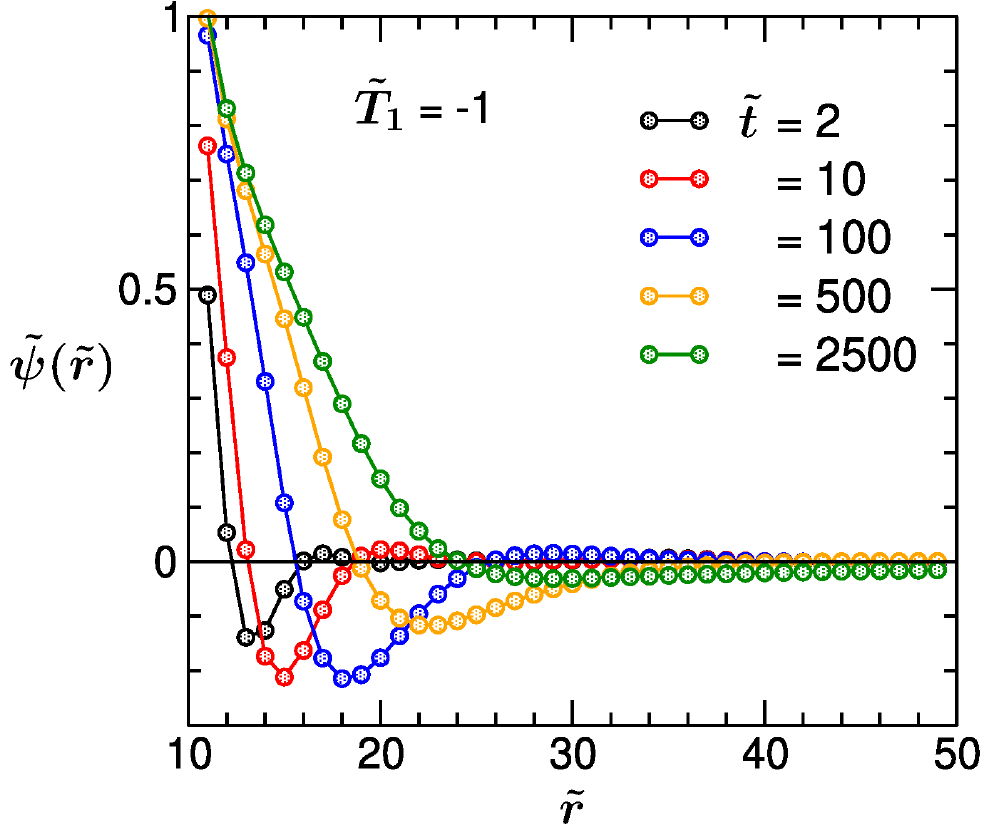}
\caption{Angularly averaged order parameter (OP) profile $\tilde{\psi}(\tilde{r})$ for the radial distance 
$\tilde{r}$ with respect to 
the single centered colloid, at five different times $\tilde{t}$, starting from the colloid's surface. 
The A phase ($\tilde{\psi}(\tilde{r}) > 0$) accumulates near the colloid surrounded by the B phase 
($\tilde{\psi}(\tilde{r}) < 0$). Away from the colloid, $\tilde{\psi}(\tilde{r})$ is around zero. 
Over time, the A phase domain broadens near the colloid and the maximum of OP 
increases while the B phase domain spreads into the bulk.}
\label{fig:1colloid_OP_T-1}
\end{figure}

For greater insight into how the order parameter field changes over time, 
the angularly averaged order parameter profile $\tilde{\psi}(\tilde{r})$ is displayed in Fig. \ref{fig:1colloid_OP_T-1}. 
Positive order parameter values correspond to the A phase whereas negative values correspond to the B phase. 
If it is zero, it corresponds to the mixed state. For early time $\tilde{t}=2$ we can already observe a layer of the A phase close to the colloid and another neighboring layer of the B phase. From $\tilde{r} \approx 18$ onwards 
$\tilde{\psi}(\tilde{r})$ stays around zero.
With increasing time the maximum value of the order parameter $\tilde{\psi}(\tilde{r})=\tilde{R}$ on the colloid's surface increases and the A phase layer as well as the B phase layer broaden. 
Note that the order parameter is a conserved quantity. That means the average value is zero at any time. Consequently, the increase of the maximum value of $\tilde{\psi}(\tilde{r})$ and the broadening of the A phase near the colloid 
result in a change of the B phase: The position of the minimum as well as the point from which $\tilde{\psi}(\tilde{r})$ stays around zero shift further away from the colloid. The B phase layer broadens into the bulk which is why the absolute minimum decreases over time. 
The steady state order parameter profile is depicted at $\tilde t=2500$ where a very wide layer of 
A phase forms and the adjacent B phase order parameter has much smaller absolute value but extends over a larger radial distance 
such that the total order parameter is zero.
\vspace{0.3cm}

\FloatBarrier
\FloatBarrier

\begin{figure}[h]
\centering
\includegraphics[scale=1]{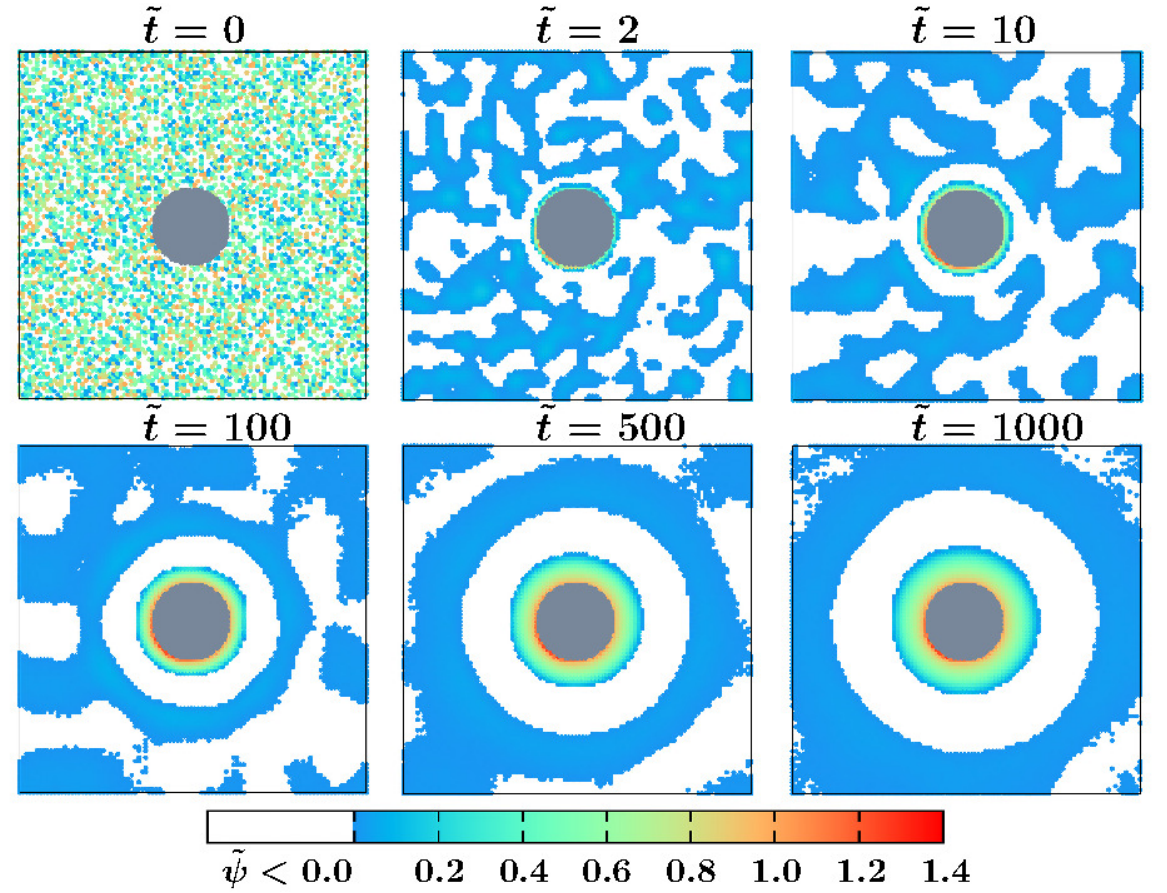}
\caption{Snapshots of the binary solvent during phase separation around a quenched colloidal particle. 
Results correspond to a deep temperature quench at $\tilde T_1=-9$. All other parameters are same as 
in Fig. \ref{fig:1colloid_snapshot_T-1}. Starting from very early times, a surface layer consisting of A phase 
($\tilde \psi(\tilde r, \tilde t)>0$) forms on the colloid's surface and spinodal-like patterns exist in the bulk. 
With time, a greater number of layers of A phase form, as compared to a small quench. The layered patterns also 
extend further away into the bulk. Note that in between two layers of A phase, a layer of B phase 
($\tilde \psi(\tilde r, \tilde t)<0$) form. }
\label{fig:1colloid_snapshot_T-9}
\end{figure}

In order to understand how the structure formation process gets influenced by the 
quench temperature, in Fig. \ref{fig:1colloid_snapshot_T-9}, the evolution snapshots are presented 
for a deeper quench ${\tilde T}_1=-9$. All other system parameters are the same as in Fig. \ref{fig:1colloid_snapshot_T-1}.
The qualitative nature of surface layer formation is same as for a small quench; at 
early time, a surface layer of phase $\tilde \psi>0$ forms and it thickens with time. 
In the bulk spinodal patterns coarsen. One striking difference in this case from the small quench is 
that at very late time $\tilde t=100$ two layers of phase $\tilde \psi>0$ form and this 
second layer gets even more prominent with increasing time. For a small quench, all the way 
upto a time when the system reaches a stationary state, only one layer forms. 

\FloatBarrier
\FloatBarrier

\begin{figure}[h]
\centering
\includegraphics[scale=1]{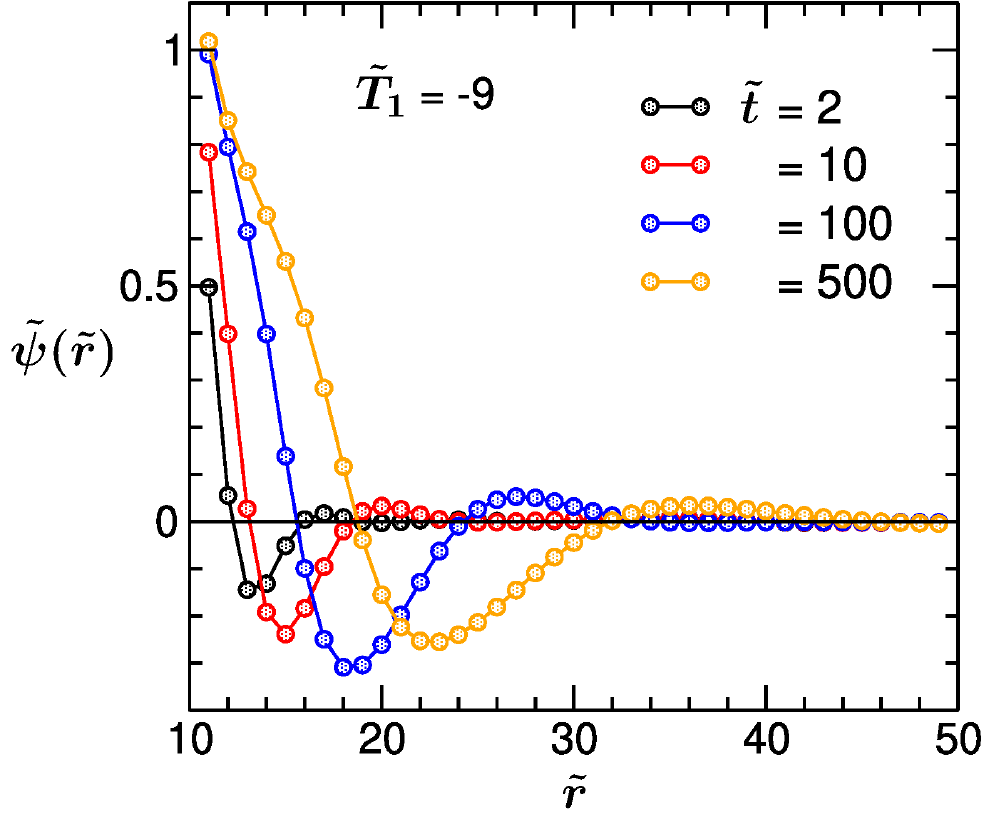}
\caption{Same as Fig. \ref{fig:1colloid_OP_T-1}, but for a deep quench $\tilde T_1=-9$. The initial 
temperature is $\tilde T_0=0.2$. All other system parameters are same as in Fig. \ref{fig:1colloid_OP_T-1}. 
Qualitatively we see the same time evolution of the order parameter $\tilde \psi(\tilde r, \tilde t)$ profile as for a small quench. A major difference is observed at late times: for a deep quench 
two prominent A phase layers are observed whereas for the small quench only one A rich layer forms. 
For deep quenches a greater number of rings form. }
\label{fig:1colloid_OP_T-9}
\end{figure}

In Fig. \ref{fig:1colloid_OP_T-9}, the angularly averaged order parameter (OP) profile $\tilde{\psi}(\tilde{r})$ 
is plotted as a function of 
the radial distance $\tilde r$ for the deep quench to $\tilde T_1=-9$. Qualitative features of 
the time evolution of the OP at early times is the same as in Fig. \ref{fig:1colloid_OP_T-1}. However, at late times 
a strong difference is observed. We see that at $\tilde t=100$, 
for a deep quench a second A-rich layer at $\tilde r\simeq 27$ 
stands out as compared to a small quench $\tilde T_1=-1$. This difference gets even stronger at a much later time $\tilde t=500$. At this time, for a small quench the system has 
almost reached the steady state and only one A phase layer is seen. Whereas, for a deep quench, 
a very thick second A phase layer is there. To understand the physical meaning behind this, 
next we look at the temperature field.

\FloatBarrier
\FloatBarrier

\begin{figure}[h]
\centering
\includegraphics[scale=1]{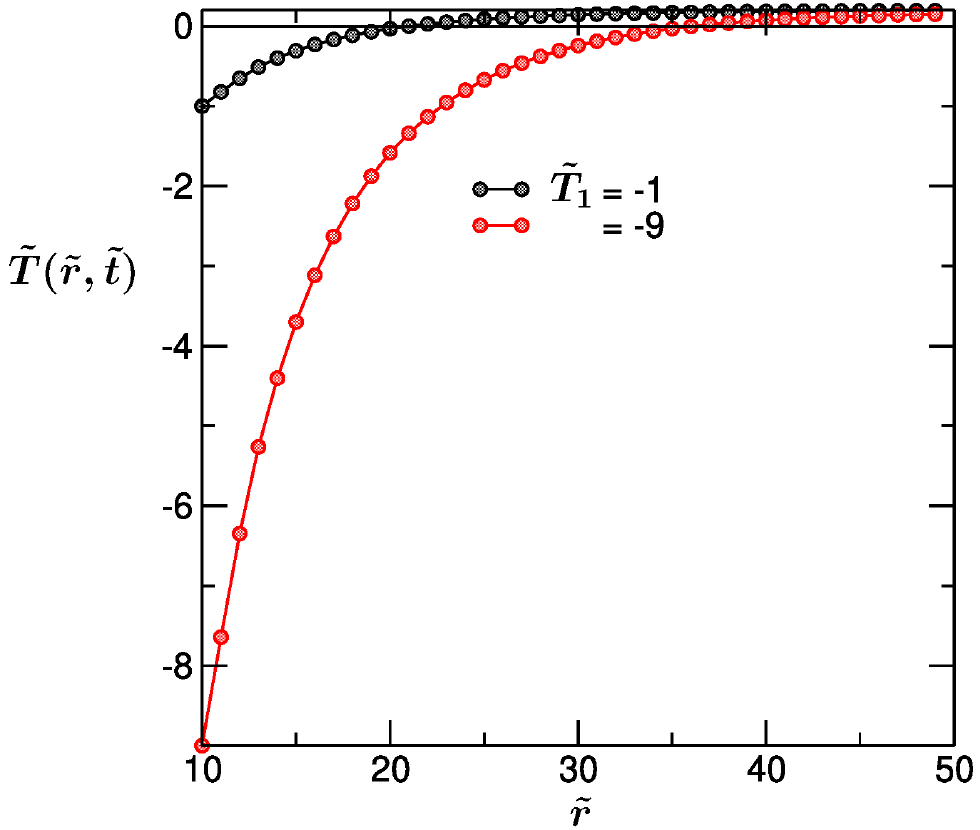}
\caption{Angularly averaged temperature profile $\tilde T(\tilde{r}, \tilde t)$ vs. the 
radial distance $\tilde{r}$ from the colloid center. Results correspond to a small ($\tilde T_1=-1$) and 
a deep ($\tilde T_1=-9$) quench. For the deep quench, the temperature field attains negative value 
($\tilde T(\tilde r) < T_c$) over a much larger distance from the colloid. This has an impact on 
the local structure formation process. }
\label{fig:1colloid_temp_compare}
\end{figure}

In Fig. \ref{fig:1colloid_temp_compare}, the angularly averaged temperature profile 
$\tilde T(\tilde{r}, \tilde t)$ is shown vs. the radial distance $\tilde r$, in the 
steady state. Results for small and deep quenches are plotted. We see that for a small quench $\tilde T(\tilde r, \tilde t)$ 
takes up negative value ($<0$) only for a smaller space region 
$\tilde r <20$. But, for a deep quench, $\tilde T(\tilde r, \tilde t)$ is negative for a much larger 
distance up to $\tilde r \simeq 38$. Phase separation occurs for negative $\tilde T$, so 
the local 
phase separation for a deep quench extends for a larger spatial region and multiple rings form.
\vspace{0.5cm}

\FloatBarrier
\FloatBarrier

We compare the coarsening patterns that we observe in this work to the patterns 
in \cite{roypre}. There, the layers of phases propagate through the entire simulation box, unlike 
the patterns observed in this thesis. This is because the stationary temperature profile in 
\cite{roypre} is $\tilde T_1=-1$ everywhere in the system. So, the 
solvent phase separates globally. On the other hand, for the b.c. used in this thesis, a temperature 
gradient always exists in the system, with the temperature value on the surface of the colloid and at the 
outer edges of the simulation box being $\tilde T_1$ and $\tilde T_0$, respectively. This
leads to a local phase separation and less number of rings form.

\section{Two colloid system}

We present results for the two-colloid system. 
In this case, two identical spherical colloids are 
kept fixed in the binary solvent. 
The system is initially at a temperature above $T_c$ and at time $\tilde t=0$, 
both colloids are quenched to a temperature $\tilde T_1$ below $T_c$ and have same 
surface properties. Both colloids are equally far away from the center of the simulation box 
and lie in the middle of the y- and z-axis. We apply the same numerical procedure as before. 

\begin{figure}[h]
\centering
\includegraphics[scale=1]{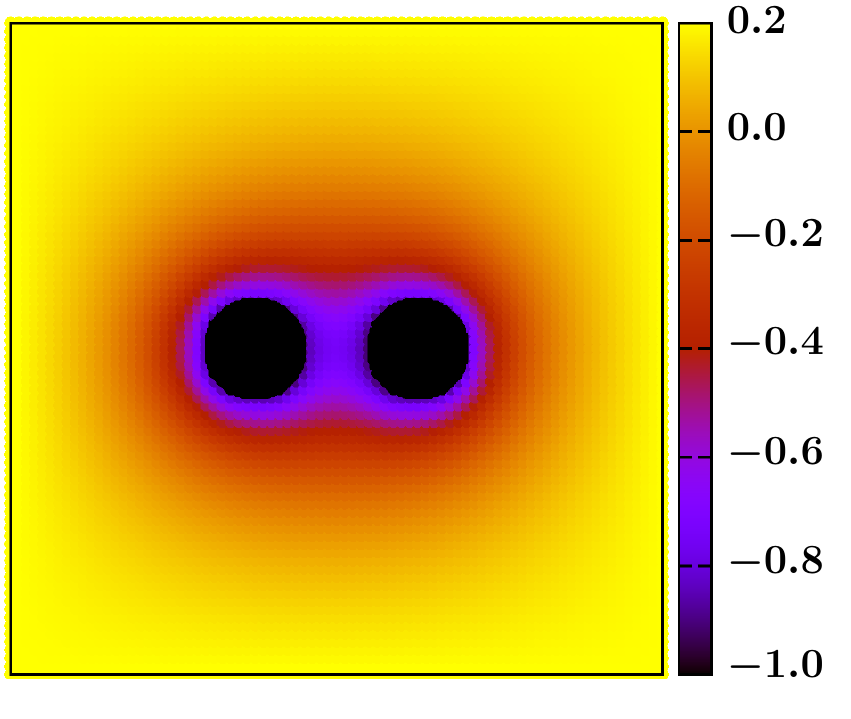}
\caption{Temperature profile of the two colloid system at an early time 
$\tilde{t}=2$ in the midplane ($z=L/2$) of the system. 
Both colloids are quenched to a temperature $\tilde T_1=-1$ and the initial temperature is $\tilde T_0=0.2$. 
The temperature at the outer boundaries of the box is kept fixed at $\tilde{T}_0=0.2$. 
A large temperature gradient is established and 
the temperature fields of the two colloids get coupled with each other.}
\label{fig:2colloids_temp}
\end{figure}

In Fig. \ref{fig:2colloids_temp}, we show the temperature profile in the midplane of the system $z=L/2$, 
at an early time $\tilde t=2$. System sizes and colloid radius that we choose are $\tilde L=80$, 
$\tilde R=5$ and the surface parameters are $\tilde \alpha=0.5$, $\tilde h=1$. 
The color bar shows different values of the temperature.
Both colloids are at temperature $-1$, marked by the black color. Near both colloids' surfaces 
a large temperature gradient is visible, 
while close to the outer edges of the simulation box values of $\tilde T(\vec r)$ are 
close to the initial temperature $\tilde{T}_1$. Note that outer edges are maintained at $\tilde T_0=0.2$ via 
suitable b.c. The purple region corresponding to a low temperature ($\tilde{T}(\vec{r},t) \approx -0.75$) forms a 
temperature bridge connecting both colloids. Thus, the temperature fields of the two colloids get coupled. 
This kind of coupling should have immediate influence in the order parameter field as well.

\FloatBarrier
\FloatBarrier

\begin{figure}[bth]
\centering
\includegraphics[scale=1]{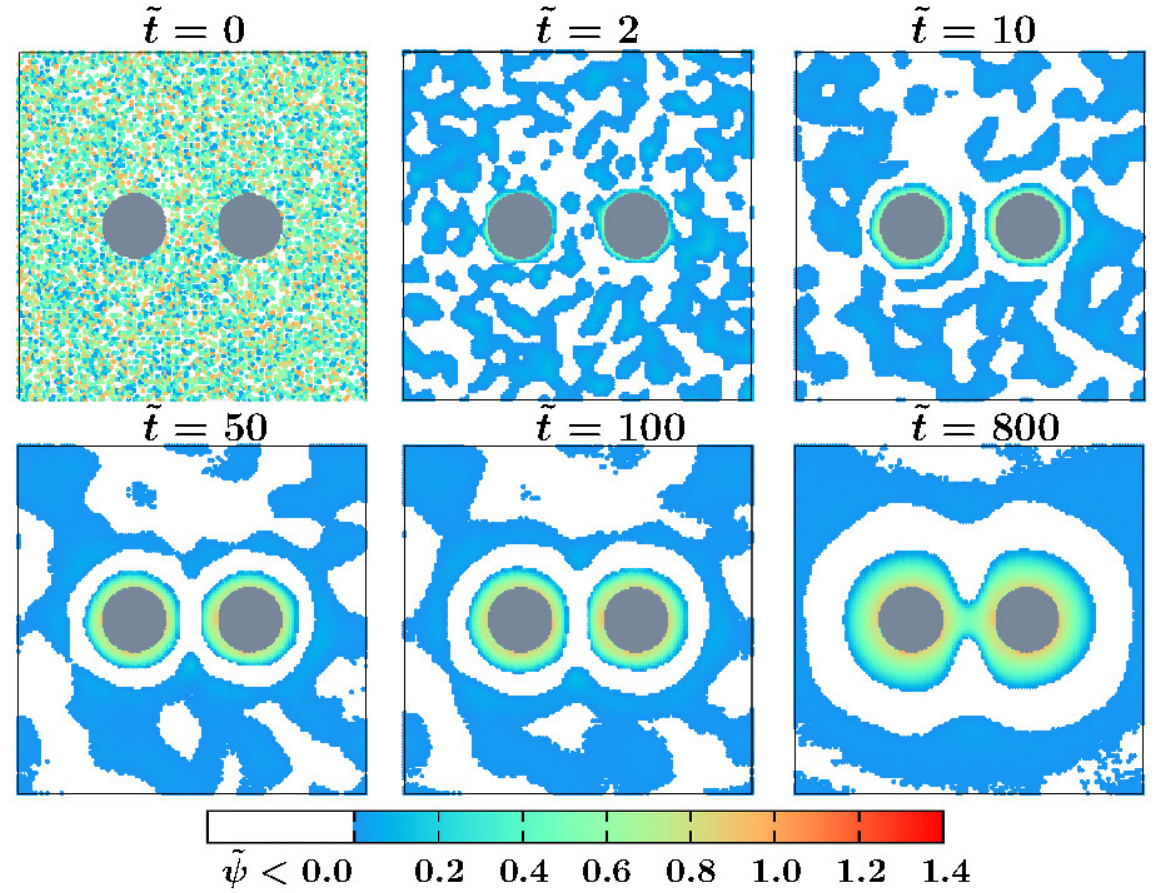}
\caption{Snapshots of temperature-gradient induced coarsening of a binary solvent 
around two spherical colloidal particles at six different times. 
Results correspond to $\tilde{L}=120$, $\tilde{R}=10$, $\tilde{\psi}_0=0$, $\tilde{T}_0=0.2$ and $\tilde{T}_1=-1$. At early time the A phase accumulates around each colloid. At $\tilde{t}=20$ the inner A phase layers of both colloid merge resulting in a coupled layer formation. In the bulk a spinodal-like pattern persists over time. }
\label{fig:2colloids_snapshot}
\end{figure}

Next, in Fig. \ref{fig:2colloids_snapshot}, we present the coarsening 
snapshots in the midplane of the system $z=L/2$ at 
different times. The system parameters are chosen to be $\tilde{L}=120$, $\tilde{R}=10$, $\tilde{\psi}_0=0$, 
$\tilde{T}_0=0.2$ and $\tilde{T}_1=-1$. At early time, the evolution around each colloid is qualitatively 
similar to what we saw for the single colloid. On each colloid a surface layer with A phase ($\tilde \psi>0$) forms 
and spinodal-like patterns exist away from the colloids. With time, these layers get thicker and the value of the order parameter 
on the surface layer increases (clear from different colors in the figure), 
while their shape stays circular.
At the same time, the spinodal domains also get bigger. At very late time ($\tilde t=800$), the surface layers of each colloid 
touch each other, they get coupled and a bridge forms. This liquid bridge is `dumbbell'-shaped. With further time, 
this bridge will evolve. But we do not present results from later times because the simulations are expensive.  

\FloatBarrier
\FloatBarrier

\begin{figure}[bth]
\centering
\includegraphics[scale=1]{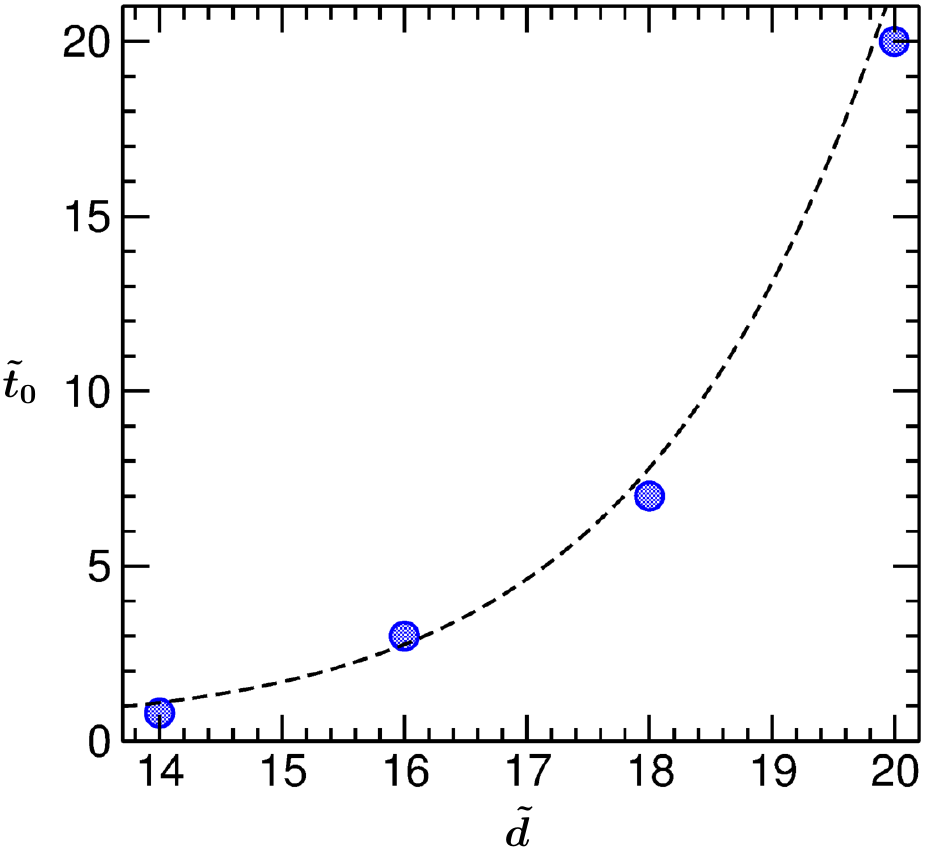}
\caption{Plot of the time $\tilde t_0$ at which the individual surface layers for two colloids merge with each other vs. the separation distance $\tilde d$ between the two colloids. 
The dashed curve is a guide to the eye.}
\label{fig:merging}
\end{figure}

We find that the time $\tilde t_0$ at which two individual layers (for each colloid) merge depend on the separation distance $\tilde d$ between the two colloids. 
In Fig. \ref{fig:merging}, this is shown for various values of $\tilde d$. The symbol there corresponds to simulation data and the dashed curve a guide to the eye. 

\FloatBarrier
\FloatBarrier

\begin{figure}[bth]
\centering
\includegraphics[scale=1]{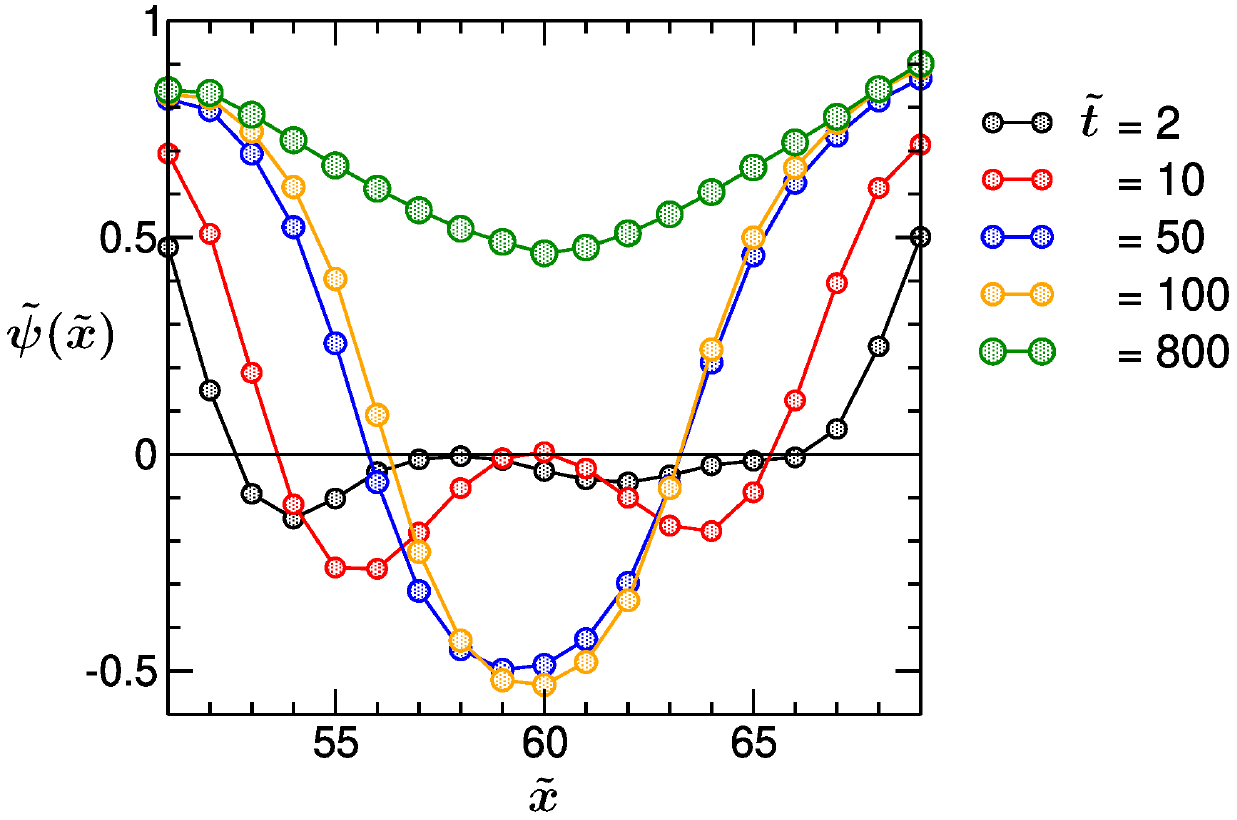}
\caption{Plot of the order parameter (OP) as a function of the $x$-coordinate in between two colloids. 
Results correspond to $\tilde L=120$, $\tilde R=10$, $\tilde T_0=0.2$, $\tilde T_1=-1$, 
$\tilde \alpha=0.5$, and $\tilde h=1$. 
$\tilde x=50$ and $\tilde x=70$ depict two colloid surfaces and the space in between corresponds to the solvent. 
The OP value along the $x$-axis in the midplane of the system is presented. A layer of phase $\tilde{\psi}>0$ and a neighboring depletion layer of $\tilde{\psi}<0$ 
form near each colloid whereas $\tilde{\psi} \approx 0$ in the middle. 
Over time the layers broaden and the maximum value of OP on the colloid surfaces increases. At $\tilde{t} = 50$ the depletion layers have merged so that one sinusoidal layer of $\tilde{\psi} < 0$ exists in the middle. At late time $\tilde{t}=800$ the layers of phase $\tilde{\psi}>0$ have broadened over the whole region, displaced the other layer and merged into one covering both colloids.}
\label{fig:2colloidsOP}
\end{figure}

In Fig. \ref{fig:2colloidsOP}, the order parameter (OP) value $\tilde \psi(\tilde x, \tilde t)$ is plotted as a function of the 
$x$-coordinate, along the $x$-axis in the midplane $z=L/2$ of the system. In this case, $\tilde L=120$, $\tilde R=10$ and the two 
colloids are placed at $(40,60,60)$ and $(80,60,60)$, respectively. The OP in the region in between the two colloids 
is presented in Fig. \ref{fig:2colloidsOP}. Note that $\tilde{x}=50$ and $\tilde{x}=70$ denote the surface of the colloids. At very early time 
$\tilde t=2$, a surface layer of the phase $\tilde \psi >0$, followed by a depletion layer of $\tilde \psi<0$ form 
on both colloids, as visible in the figure. In the middle the OP stays close to zero. With increasing time, thickness of 
these surface layers as well as the depletion layers increase and the region with zero OP almost vanishes. 
At time $\tilde t=50$ onwards, the two depletion layers have merged and only one sinusoidal layer of phase 
$\psi <0$ exists in between the two colloids. The surface layers have also increased in thickness. Finally, at very late time 
$\tilde t=800$, the OP value in the whole region in between the two colloids has become positive. This corresponds to a 
physical situation in which a liquid bridge of phase $\tilde \psi>0$ has formed covering both colloids. This total OP 
in the system, however, stays zero which is clear from the snapshots in Fig. \ref{fig:2colloids_snapshot}. 
Note that the profiles in Fig. \ref{fig:2colloidsOP} will be symmetric with respect to both colloids upon averaging over multiple initial 
configurations. Here we present them only from a single realization. 

\FloatBarrier
\FloatBarrier

\section{Force acting on the colloids}

We also investigate the forces acting on two suspended colloidal particles in a near-critical solvent. 
This force arises due to the gradient of the chemical potential of the solvent. During the 
non-equilibrium coarsening processes, the temperature field and the coupled order parameter field 
change with time. As a result, the chemical potential field also changes with time. 
The generalized force $\tilde F$, which we are interested in, is proportional to 
$\tilde \psi(\tilde r, \tilde t) \nabla \tilde \mu(\tilde r, \tilde t))$. The net force acting on a colloid is 
given by the integral over the colloid surface. For a single colloid, this force $\tilde F$ is expected to be zero and 
the colloid does not move. However, when there are two colloids in the solvent very close to each other 
(i.e., the separation distance is small enough) a non-zero force may act on each colloid.

\begin{figure}[bth]
\centering
\includegraphics[scale=1]{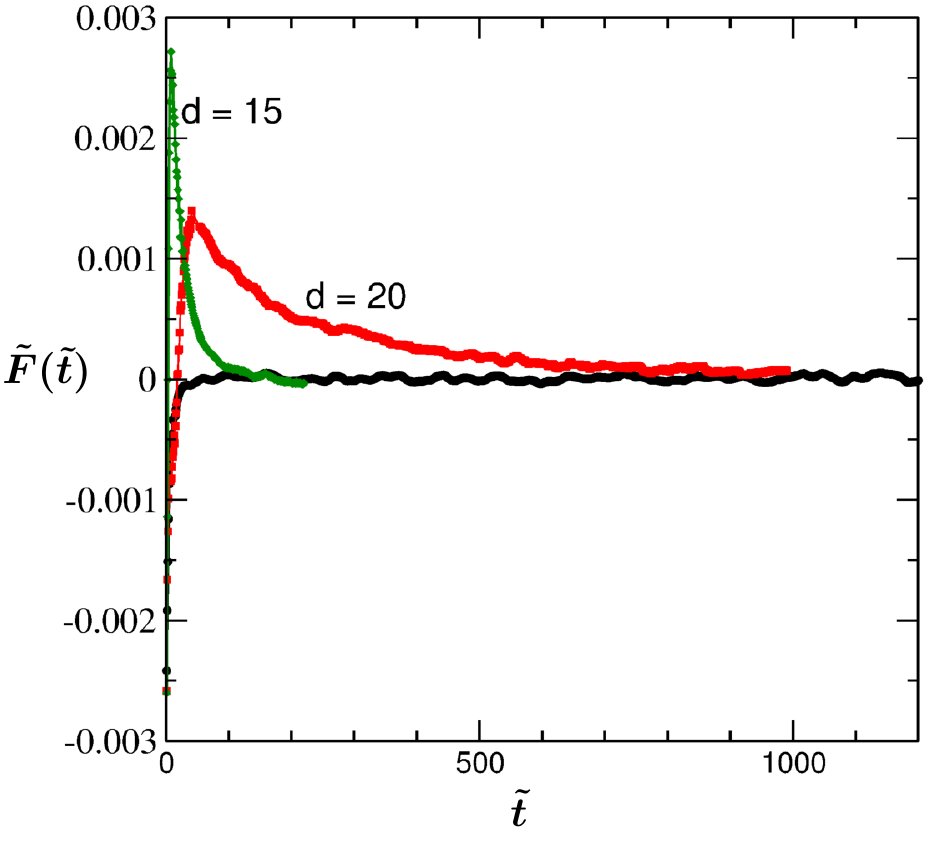}
\caption{Plot of the force $\tilde F(\tilde t)$ acting on one of the two colloids during coarsening, resulting from a chemical potential gradient as a function of time $\tilde{t}$. See main text for the definition of $\tilde F(\tilde t)$. The maximum value of $\tilde F(\tilde{t})$ is larger for shorter separation distances.}
\label{fig:force1}
\end{figure}

In Fig. \ref{fig:force1}, we plot the $x$-coordinate of the averaged force $\tilde F$ as a function of time $\tilde t$ following a 
temperature quench of both colloids from an initial temperature $\tilde T_0=0.2$ to $\tilde T_1=-1$.  
The system size and the colloid radius considered are $\tilde L_x=140$, $\tilde L_y=\tilde L_z=50$, 
$\tilde R=5$. The first and second colloids are kept at locations $(65, 25, 25)$, $(85, 25, 25)$, respectively. 
That means we have a center to center separation distance $\tilde d=20$. The data is averaged over 5 independent 
initial conditions. The black data corresponds to the force acting on the single colloid.
$\mathcal D$ is taken to be $100$. Here, we plot only the force acting on the first colloid. 
As a reference, the force acting on a single colloid is also 
presented in this plot. For the single colloid we take $\tilde L_x=100$, $\tilde L_y=\tilde L_z=50$, 
$\tilde R=5$ and the colloid is placed fixed at the center of the box. 
All other parameters for the single colloid case are the same as for the two-colloid case. 
First of all, following a very early-time window, the average force $\tilde F$ for a single colloid 
saturates to zero very quickly, i.e., the colloid can not move. For the two-colloid case, following a 
temperature quench the colloid experiences a non-zero force $\tilde F$ which initially increases with 
increasing time, then it reaches a certain threshold and after that it 
starts decaying and at very late time finally 
saturates to zero. The time when this force decays to zero depends on the separation distance 
$d$ between the two colloids. It decays faster for smaller $\tilde d$. The maximum value of the force $\tilde F$ is 
also larger for smaller $\tilde d$. 

Since this force $\tilde F$ appears only upon confinement, let us denote the time-dependent force 
acting on colloid 1 for the two-colloid system as $\tilde F_\text{two}(\tilde t)$ and the force acting on a single 
colloid as $\tilde F_\text{single}$. One can then write, $\tilde F_\text{two}(\tilde t)=\tilde F_\text{single}+ \tilde F_\text{excess}$. As clear from Fig. \ref{fig:force1}, this excess force $\tilde F_\text{excess}$ starts from zero and decays to zero at very late time. In the intermediate time this is non-monotonic. In future, we will do a scaling analysis of this excess force $\tilde F_\text{excess}$.

\FloatBarrier
\FloatBarrier
\chapter{Summary and outlook}\label{ch:summary}

We have presented results for the numerical simulation of 
the non-equilibrium dynamics of the temperature-gradient induced 
coarsening of a binary solvent around spherical colloidal particles 
at its critical concentration. Initially, the colloid as well as 
the solvent are set to temperature $\tilde T_0$ above the critical 
temperature of the solvent. Then, the colloid is quenched to a 
temperature $\tilde T_1$ below the critical temperature, establishing 
a temperature gradient in the system. The temperature gradient is 
maintained by suitable boundary condition (b.c.). The outer boundaries 
of the simulation box are always at the initial temperature $\tilde T_0$ 
and the colloid is always at the quench temperature $\tilde T_1$. 
We use a b.c. that is different from the ones used in earlier works 
\cite{roypre, roysoftmatter}. We simulated an one colloid system 
as well as a two colloid system and incorporated the colloid's adsorption 
preference by applying the \textbf{Robin} boundary condition. The observed 
quantities are the time-dependent temperature field and the time-dependent 
order parameter field which are described by the modified Cahn-Hilliard-Cook 
equation coupled with the heat diffusion equation. 

We found that circular layers of phase A and phase B, referring to the 
components of the solvent, form around the colloid. Two neighboring 
layers are of opposite phase. Away from the surface, spinodal-like patterns 
exist. The layers near the colloid evolve and broaden into the bulk while 
increasing their respective concentration ($c_A$ for phase A and $c_B$ for 
phase B) over time. By simulating the two colloid system we observed the 
coupling of the layer formation of each colloid. At very early time one 
layer forms close to each colloid separately. As time progresses, these 
layers merge into a `dumbbell'-shaped liquid bridge which connects both 
colloids. For deep quenches we observed the formation of a greater number 
of layers. We also studied the force acting on 
suspended colloids due to the gradient in the chemical potential field 
close to the surface. The force acting on a single colloid is zero which 
means the colloid does not move. For two colloids suspended in the 
near-critical solvent, for short separation distances, a non-zero force 
acts on each colloid. This force increases in magnitude with decreasing 
separation distance. 

In our model, we assumed the mobility $M$ and the thermal diffusivity $D_{th}$ 
to be constant; although in reality they depend on $\psi$. This leaves 
space for an improvement in future works. The results in this thesis 
are obtained for colloids in the bulk. It is certainly interesting to 
simulate our system also in confinement and calculate the force acting 
on the colloids in confinement. In general, our model offers a lot of 
possibilities for simulations to investigate influences on the coarsening 
patterns and temperature profile. It is possible to simulate more colloids 
in one system, change their adsorption preferences and their positioning. 
Also, Janus particles can be used where the two hemispheres of the particle 
have different temperature and different surface properties. Additionally, 
temperature quenches can be activated in a sinusoidal way.

\appendix
\cleardoublepage
\chapter{Appendix}\label{ch:appendix}

\section{Non-dimensionalization of Cahn-Hilliard-Cook and heat diffusion equations} \label{sec:nondimensionalizationCHCandHD}

In order to convert the dimensional equations Eqs. (\ref{eq:CHCnoise}) and 
(\ref{eq:HDdim}) to corresponding dimensionless forms, let us apply 
the substitutions $\vec{r} = \tilde{\vec{r}}\vec{r}_0$, $t = \tilde{t}t_0$, 
$\psi(\vec{r},t) = \tilde{\psi}(\tilde{\vec{r}},\tilde{t})\psi_0$, 
 and $\eta(\vec{r},t) = \tilde{\eta}(\tilde{\vec{r}},\tilde{t})\eta_0$. 
The quantities with index $0$ are the rescaling coefficients and the symbol 
\textasciitilde \ above the quantities refers to the quantities without dimension.

As a preparation, the derivative with respect to t and the derivative with 
respect to r are converted as:
\begin{align}
\frac{\partial }{\partial r} &= \frac{\partial }{\partial \tilde{r}}\frac{\partial \tilde{r} }{\partial r} = \frac{1}{r_0} \frac{\partial}{\partial \tilde{r}}, \label{eq:nablaentdim} \\
\frac{\partial }{\partial t} &= \frac{1}{t_0} \frac{\partial}{\partial \tilde{t}}.
\end{align}

For the sake of completeness, let us recall here again the dimensional form 
of the CHC equation in Eq. (\ref{eq:CHCnoise})
\begin{align} \label{eq:CHCdim-2}
\frac{\partial \psi(\vec{r},t)}{\partial t} 
&= \frac{M k_B T_c}{v}~ \nabla^2 \left[ \tilde{T}(\vec{r},t) \psi(\vec{r},t) +u\psi(\vec{r},t)^3 \right. \nonumber \\ 
&\quad \left. - C \nabla^2 \psi(\vec{r},t) \right] +\eta(\vec{r},t)
\end{align}

Inserting the above-mentioned substitutions and the converted 
derivations in Eq. (\ref{eq:CHCdim-2}) yields 

\begin{align} \label{eq:CHCdimensionlessprocess}
\frac{\partial \tilde{\psi}(\tilde{\vec{r}},\tilde{t})}{\partial \tilde{t}} &=  \frac{M k_B T_c}{v} \frac{t_0}{r_0^2 } \tilde{\nabla}^2 \left[ \tilde{T}(\tilde{\vec{r}},\tilde{t}) \tilde{\psi}(\tilde{\vec{r}},\tilde{t}) +u \psi_0^2 \tilde{\psi}(\tilde{\vec{r}},\tilde{t})^3 \right. \nonumber \\
&\quad  - \frac{C}{r_0^2} \left. (\tilde{\nabla}^2\tilde{\psi}(\tilde{\vec{r}},\tilde{t}))  \right] + \frac{t_0}{\psi_0}\eta_0 \tilde{\eta}(\tilde{\vec{r}},\tilde{t}).
\end{align}

Analogously this is done for the heat diffusion equation (\ref{eq:HDdim}):

\begin{align} \label{eq:HDdimensionlessprocess}
\frac{\partial \tilde{T}(\tilde{\vec{r}},\tilde{t})}{\partial \tilde{t}} = D_\text{th} \frac{t_0}{r_0^2} \tilde{\nabla}^2 \tilde{T}(\tilde{\vec{r}},\tilde{t}).
\end{align}

For equation (\ref{eq:CHCdimensionlessprocess}) to be dimensionless the 
rescaling coefficients can be chosen such that the following is satisfied:

\begin{align} \label{eq:dimensionlesscondition}
\frac{M k_B T_c t_0 |\tilde{T}_1|}{\nu r_0^2} = \frac{M k_B T_c t_0 u \psi_0^2}{\nu r_0^2} = \frac{M k_B T_c t_0 C}{\nu r_0^4} = \frac{t_0 \eta_0}{\psi_0} = 1.
\end{align}

As a result one obtains the following rescaling factors:

\begin{align}
\psi_0 &= \left(\frac{|\tilde{T}_1|}{u}\right)^{1/2}, \label{eq:rescalingcoefficientPsi0} \\
r_0 &= \left(\frac{C}{u \psi_0^2}\right)^{1/2} = \left(\frac{C}{|\tilde{T}_1|}\right)^{1/2} \label{eq:rescalingcoefficientr0} \\
t_0 &= \frac{\nu r_0^2}{M k_B T_c |\tilde{T}_1|} = \frac{\nu C}{M k_B T_c |\tilde{T}_1|^2} \text{ and} \\
\eta_0 &= \frac{\psi_0}{t_0} = \left(\frac{|\tilde{T}_1|}{u}\right)^{1/2}/\frac{\nu C}{M k_B T_c |\tilde{T}_1|^2}.
\end{align}

Using these factors in Eq. (\ref{eq:HDdimensionlessprocess}) leads to 

\begin{align}
\frac{\partial \tilde{T}(\tilde{\vec{r}},\tilde{t})}{\partial \tilde{t}} &= D_\text{th}  \frac{\nu}{M k_B T_c |\tilde{T}_1|} \tilde{\nabla}^2 \tilde{T}(\tilde{\vec{r}},\tilde{t}) \\
&= \frac{D_\text{th}}{D_\text{m} |\tilde{T}_1|} \tilde{\nabla}^2 \tilde{T}(\tilde{\vec{r}},\tilde{t}) \\
&= \mathcal{D}\tilde{\nabla}^2 \tilde{T}(\tilde{\vec{r}},\tilde{t})
\end{align}

where $D_\text{m}(T_c) = M k_B T_c / v$ is the interdiffusion constant 
of the binary solvent at temperature $T_c$ and 
$\mathcal{D} = D_\text{th}/(D_\text{m} |\tilde{T}_1|)$.
\vspace{1cm}
\section{Non-dimensionalization of the Robin boundary condition} \label{sec:nondimensionalizationRobin}
For converting the Robin boundary condition
\begin{align} \label{eq:robindim}
h = \left. \left[ \hat{\vec{n}} \cdot \nabla \psi(\vec{r},t) + \alpha \psi(\vec{r},t) \right]\right|_{\mathcal{S}}
\end{align}
to a dimensionless equation, the substitutions $\alpha = \tilde{\alpha}\alpha_0$, $\psi(\vec{r},t) = \tilde{\psi}(\tilde{\vec{r}},\tilde{t})\psi_0$, $\vec{r} = \tilde{\vec{r}}\vec{r}_0$ and $h = \tilde{h}h_0$ are used. \\
Inserting the substitutions in Eq. (\ref{eq:robindim}) and using Eq. (\ref{eq:nablaentdim}), one obtains 

\begin{align} \label{eq:robindimensionlessprocess}
\tilde{h}h_0 &= \left. \left[\frac{\psi_0}{r_0} \frac{\partial \tilde{\psi}(\tilde{\vec{r}},\tilde{t})}{\partial \tilde{r}} + \alpha_0 \psi_0 \tilde{\alpha} \tilde{\psi}(\tilde{\vec{r}},\tilde{t}) \right]\right|_{\mathcal{S}}.
\end{align}

For Eq. (\ref{eq:robindimensionlessprocess}) to be 
dimensionless the rescaling coefficients are chosen analogously to 
Sec. \ref{sec:nondimensionalizationCHCandHD} so that
\begin{align}
\frac{\psi_0}{r_0 h_0} = \frac{\alpha_0 \psi_0}{h_0} = 1.
\end{align}
As a result one obtains the following rescaling factors: 
\begin{align}
\alpha_0 &= \frac{1}{r_0} \overset{\text{(\ref{eq:rescalingcoefficientr0})}}{=} \left( \frac{|\tilde{T}_1|}{C} \right) ^{1/2} \\
h_0 &= \frac{\psi_0}{r_0} \underset{\text{(\ref{eq:rescalingcoefficientr0})}}{\overset{\text{(\ref{eq:rescalingcoefficientPsi0})}}{=}} \frac{|\tilde{T}_1|}{(uC)^{1/2}}.
\end{align}
\cleardoublepage%
\defbibheading{bibintoc}[\bibname]{%
  \phantomsection
  \manualmark
  \markboth{\spacedlowsmallcaps{#1}}{\spacedlowsmallcaps{#1}}%
  \addtocontents{toc}{\protect\vspace{\beforebibskip}}%
  \addcontentsline{toc}{chapter}{\tocEntry{#1}}%
  \chapter*{#1}%
}
\printbibliography[heading=bibintoc]

\cleardoublepage%
\pdfbookmark[0]{Declaration}{declaration}
\chapter*{Declaration}
\thispagestyle{empty}
\begin{otherlanguage}{ngerman}
Ich erkl"are hiermit,
\begin{itemize}
\item dass ich diese Bachelorarbeit selbst"andig verfasst habe,
\item dass ich keine anderen als die angegebenen Quellen verwendet und alle w"ortlich oder sinngem"a"s aus anderen Werken "ubernommenen Aussagen als solche gekennzeichnet habe,
\item dass die eingereichte Arbeit weder vollst"andig noch in wesentlichen Teilen Gegenstand eines anderen Pr"ufungsverfahrens ist,
\item dass ich die Arbeit weder vollst"andig noch in Teilen bereits ver"offentlicht habe
\item und dass der Inhalt des elektronischen Exemplares mit dem des Druckexemplares "ubereinstimmt.
\end{itemize}
\end{otherlanguage}
\bigskip

\noindent\textit{\myLocation, \myTime}

\smallskip

\begin{flushright}
    \begin{tabular}{m{5cm}}
        \\ \hline
        \centering\myName \\
    \end{tabular}
\end{flushright}

\cleardoublepage\pagestyle{empty}

\hfill

\vfill

\pdfbookmark[0]{Colophon}{colophon}
\section*{Colophon}
This document was typeset using the typographical look-and-feel \texttt{classicthesis} developed by Andr\'e Miede and Ivo Pletikosić.
The style was inspired by Robert Bringhurst's seminal book on typography ``\emph{The Elements of Typographic Style}''.

\bigskip

\end{document}